\documentclass[12pt,a4paper]{article}

\usepackage{amssymb}
\usepackage{setspace}
\usepackage{indentfirst}

\usepackage{graphicx}

\begin{document}

\begin{titlepage}
\begin{center}

\vspace*{25mm}

\begin{spacing}{1.7}
{\LARGE\bf On CMB B-modes and the Onset of Inflation}
\end{spacing}

\vspace*{25mm}

{\large
Noriaki Kitazawa
}
\vspace{10mm}

Department of Physics, Tokyo Metropolitan University,\\
Hachioji, Tokyo 192-0397, Japan\\
e-mail: noriaki.kitazawa@tmu.ac.jp

\vspace*{25mm}

\begin{abstract}
The temperature perturbations of the cosmic microwave background radiation (CMB)
 appear systematically suppressed, at large angular scales,
 with respect to the prediction of the $\Lambda$CDM concordance model.
This behavior might be a glimpse of the inflaton deceleration from fast-roll at the start of inflation.
If this were true,
 the spectrum of primordial tensor perturbations should be distorted in a corresponding way,
 with a suppression of large angular scale B-mode polarization
 that might be within reach for future probes like LiteBIRD.
We investigate this scenario by a semi-analytic method
 based on a truncation of the B-mode polarization spectrum,
 which in our opinion has the virtue of highlighting the underlying Physics.
We present some mild evidence that, with sufficient knowledge of the reionization process,
 the suppression of large-scale B-modes might be within reach.
\end{abstract}

\end{center}
\end{titlepage}

\doublespacing

\section{Introduction}
\label{sec:introduction}

The temperature perturbations and the polarizations of the cosmic microwave background radiation (CMB)
 around the $2.7$K black-body profile encode invaluable information on the Universe at its early stages.
In particular,
 they provide constraints on the dynamics of inflation,
 on the nature of dark matter and on the process of recombination.
The $\Lambda$CDM concordance model of Cosmology accounts well for CMB perturbations and polarizations,
 but also for other large-scale aspects of the Universe,
 including the present abundance of light nuclei (nucleosynthesis)
 and the present phase of acceleration~\cite{Akrami:2018vks}.

One can extract a lot of important information
 from small angular scale perturbations and polarizations of the CMB.
Yet, large angular scales contain pristine information on earlier times in the history of the Universe,
 although lower statistics and cosmic variance result in larger errors.
A systematic suppression of temperature perturbations for multipole moments $\ell<30$,
 corresponding to angular separations $\Delta\theta \simeq \pi/\ell$,
 has emerged independently in COBE, WMAP and PLANCK data
 \cite{Hinshaw:1996ut,Spergel:2003cb,Sarkar:2010yj,Gruppuso:2013dba,Ade:2013nlj,Akrami:2018odb}.
This $\Lambda$CDM model driven by high-$\ell$ data predicts low-$\ell$ temperature perturbations
 that would be larger than the observed ones, so that this lack of power at low-$\ell$ is one of the so--called low-$\ell$ anomalies.
Although the statistical significance of this low-$\ell$ suppression is not very high,
 it may be confronting us with deep hints on phenomena that occurred at the early stages of inflation
 (see \cite{Gruppuso:2015zia} for a short review).

The $\Lambda$CDM model
 rests on the Chibisov--Mukhanov~\cite{chib_mukh} primordial power spectrum of scalar perturbations,
\begin{equation}
P_S(k) \ = \ A_S \, \left(\frac{k}{k_0}\right)^{n_S-1} \ , \label{chibmukh}
\end{equation}
 where $k$ is the wave number of the perturbation,
 $k_0$ is a pivot scale, $A_S$ is the amplitude at the pivot scale
 and $n_S$ is the spectral index.
The standard slow-roll inflation yields
 almost scale-invariant primordial power spectra,
 with a red tilt consistent with the value $n_S = 0.965 \pm 0.004$
 measured by PLANCK~\cite{Aghanim:2018eyx}.
Since inflation must have ended,
 leaving way to a hot Universe by reheating,
 many phenomenological inflation models provide cut-offs in $P_S(k)$ at larger $k$.
On the other hand,
 a cut-off (or some detailed structure) of $P_S(k)$ for small values of $k$
 could in principle account for the low-$\ell$ suppression,
 if the present horizon scale were exposing us somehow,
 by a remarkable cosmic coincidence,
 to perturbations reflecting the start of inflation.
In the simplest models,
 inflation starts when a special scalar field, the inflaton,
 undergoes a transition from fast-roll to a slow-roll behavior driven by an almost flat potential
 (see \cite{Dimopoulos:2016yep} for example).
The resulting deformations of primordial power spectra
 entail a typical cut-off in $P_S(k)$, which is to approach a $k^3$--profile at small $k$
 (a typical example is discussed in detail in \cite{Destri:2009hn}).
The power cut is possibly accompanied
 by some signs of peculiar dynamics \cite{Dudas:2012vv,Kitazawa:2014dya}.
The simple deformed power spectrum
\begin{equation}
 P_S(k) \ =\  A_S\,\frac{ \left(\frac{k}{k_0}\right)^3}{\left[\left(\frac{k}{k_0}\right)^2 \ + \ \left(\frac{\Delta}{k_0}\right)^2\right]^{2-\frac{n_S}{2}}}
\label{primordial-scalar-Delta}
\end{equation}
 captures the gross features of the transition region,
 in a model independent way~\cite{Dudas:2012vv,Kitazawa:2014dya}, and
 the analysis in \cite{Gruppuso:2015xqa,Gruppuso:2017nap} showed that the value
\begin{equation}
\Delta\ =\ (0.351 \pm 0.114) \times 10^{-3} \ \ \mathrm{Mpc}^{-1}
\end{equation}
 fits reasonably well PLANCK data at large galactic latitudes.
The determination of $\Delta$ improves indeed with galactic masks
 that are larger than the minimal PLANCK standard one,
 and the value that we refer to here
 with the best statistical significance
 was obtained with a galactic mask blindly extended by about 30 degrees
 \cite{Gruppuso:2015xqa,Gruppuso:2017nap}.

The power spectrum in eq.~(\ref{primordial-scalar-Delta})
 is part of a family of analytical power spectra introduced in~\cite{Dudas:2012vv,Kitazawa:2014dya}
 to account for the transition of the Mukhanov--Sasaki potential
 from the negative values typical of fast--roll to the positive Coulomb--like barrier of slow--roll.
A lack of power along these lines emerged,
 at weak string coupling and thus somewhat reliably, from ``brane supersymmetry breaking'',
 a high-scale mechanism \cite{bsb} that presents itself in String Theory \cite{stringtheory}
 (for reviews see \cite{Angelantonj:2002ct}).
In the resulting scenario,
 a scalar is forced to emerge from an initial singularity climbing up a steep exponential potential,
 and its pre-inflationary deceleration provides a natural origin
 for the lack of power at small $k$,
 which is typically accompanied, in the resulting scenarios, by pre--inflationary peaks.
As in~\cite{Gruppuso:2015xqa,Gruppuso:2017nap},
 here we proceed from a phenomenological vantage point,
 but this scenario clearly lies behind our motivations.
This ``climbing scenario'' actually translates into a family of power spectra
 that depend on the initial conditions for the inflaton,
 as described in~\cite{Dudas:2012vv,Kitazawa:2014dya}.
Still, the simple spectrum of eq.(\ref{primordial-scalar-Delta})
 captures their generic behavior in the transition region.

The primordial tensor power spectrum
 is also affected by the transition from fast--roll to slow--roll,
 in such a way that, locally, the tensor-to-scalar ratio $r_{TS}$
 can grow by about one order of magnitude in the region \cite{Gruppuso:2017nap}.
However, in the spirit of the preceding considerations,
 here we shall ignore this subtlety and, in order to capture the gross features of the phenomenon,
 we shall resort to a similar parametrization,
\begin{equation}
 P_T(k) \ =\  A_T\,\frac{ \left(\frac{k}{k_0}\right)^3}{\left[\left(\frac{k}{k_0}\right)^2 \ + \ \left(\frac{\Delta}{k_0}\right)^2\right]^{\frac{3- n_T}{2}}}
\label{primordial-tensor-Delta}
\end{equation}
 with the same scale $\Delta$ but with a different amplitude $A_T$
 and involving the different tensor spectral index $n_T$, for which we abide to the standard notation.
It is known that
 the amplitude of tensor perturbation is small, with a tensor-to-scalar ratio $r_{TS}$
 here characterized by $r_{TS} = A_T/A_S < 0.07$ \cite{Aghanim:2018eyx}, so that
 one can not see the effect in correlations between temperature perturbations and E-mode polarizations,
 where scalar perturbations dominate.
On the other hand, in correlations of B-mode polarizations,
 namely $C^{\rm BB}_\ell$ or $D^{\rm BB}_\ell = (\ell(\ell+1)/2\pi) C^{\rm BB}_\ell$,
 the contributions of tensor perturbations dominate over those of scalar perturbations,
 especially in low-$\ell$ region.
Matters are indeed complicated by the fact that, in the region of $\ell \gtrsim 10$,
 large contributions to B-mode polarizations of scalar perturbations induced by gravitational lensing
 dominate over those from primordial tensor perturbations \cite{Zaldarriaga:1998ar}.
As a result,
 the distortion of the primordial tensor perturbations of eq.(\ref{primordial-tensor-Delta})
 can directly reflect on $D^{\rm BB}_\ell$ only in the region of $\ell \lesssim 10$.
The extent of this effect
 depends on the actual value of tensor-to-scalar ratio and the actual magnitude of the lensing effect.
In this paper we work with $r_{TS}=0.03$ and assume that
 the lensing effect is negligible, or subtracted somehow by some theoretical and observational estimates.

The primordial CMB radiation is not expected to be polarized,
 and polarization is induced via the process of Thomson scattering with free electrons.
The free electrons needed for the polarization at small angular scales
 are produced in the process of recombination, which did not perfectly neutralize the Universe,
 so that a certain amount of electrons and ions remained available to affect the CMB polarization.
The density of these free electrons was diluted by the expansion of the universe,
 while the free electrons inducing polarization at larger angular scale perturbations,
 $\ell \lesssim 10$, were produced in the process of reionization.
Therefore,
 a proper knowledge of the reionization process
 impinges on this route to attain further evidence for the start of inflation.

In this work we investigate the B-mode polarization power spectrum $D^{\rm BB}_\ell$,
 concentrating on the region of $\ell \lesssim 10$.
We use the semi-analytic method pioneered by Polnarev \cite{Polnarev:1985},
 with a truncation that clearly highlights the underlying Physics.
Comparing predictions
 resting on the power spectrum of eq.~(\ref{primordial-tensor-Delta}) with those of $\Lambda$CDM model,
 we provide some evidence to the effect that cosmic-variance limited measurements by LiteBIRD,
 for example, might be able to provide further indirect evidence for these types of manifestations
 of the start of inflation.
In the next section
 we describe our semi--analytic estimate of the B-mode polarization power spectrum $D^{\rm BB}_\ell$.
To this end,
 we introduce a convenient truncation, at the price of sacrificing quantitative precision.
This simplified procedure is possible, since we concentrate on the low-$\ell$ region,
 and the period of time evolution is merely from typical red shifts
 corresponding to the start of reionization $z_{\rm ion} \simeq 8$ to the present epoch.
In section \ref{sec:results}
 we present B-mode polarization power spectra,
 taking into account several possible decorations of the primordial tensor power spectrum of
 eq.~(\ref{primordial-tensor-Delta}).
We also investigate how details of the reionization process can affect the results.
To this end,
 we consider the effects on the value of $z_{\rm ion}$
 and of the difference between instantaneous reionization and slow increase of the free electron density,
 while keeping the optical depth constant at the value $\tau \simeq 0.054$ \cite{Aghanim:2018eyx}.
In the last section
 we conclude with a discussion of the possible statistical significance
 of this type of insights on the start of inflation.

\section{The Semi--Analytic Strategy}
\label{sec:strategy}

We follow the formalism pioneered by Polnarev in \cite{Polnarev:1985},
 where he introduced the Boltzmann equation for an array of occupation numbers of polarized radiation.
The formalism is a counterpart of the semi-analytic formalism for temperature perturbations
 pioneered by Mukhanov \cite{Mukhanov:2003xr,Mukhanov:2005sc}.
In the following we are going to use the formalism without providing a detailed review,
 which the reader can find in~\cite{Cabella:2004mk}
 and in the review sections of~\cite{Keating:1997cv,Pritchard:2004qp,Zhang:2005nv}.

The Boltzmann equation
 for the time evolution of photon occupation numbers encoding the evolution of polarization
 as they propagate in the background of tensor perturbations and free electrons is
\begin{equation}
 \frac{\partial\tilde{f}}{\partial\eta}
\  - \ \hat{n}_i \frac{\partial\tilde{f}}{\partial x^i}
 \ - \ \frac{1}{2} \frac{\partial h_{ij}}{\partial \eta} \hat{n}_i \hat{n}_j
   \nu \frac{\partial\tilde{f}}{\partial\nu}
 \ =\  - \ g(\eta) ( \tilde{f} - \tilde{J} )\ ,
\label{eq-rad-transfer}
\end{equation}
 where $\eta$ is the conformal time of the metric
\begin{equation}
 ds^2 = a^2 \left( d\eta^2 - ( \delta_{ij} + h_{ij}) dx^i dx^j \right)
\end{equation}
 with tensor perturbations $h_{ij}$,
 $\hat{n} = (\sin\theta\cos\phi, \sin\theta\sin\phi, \cos\theta)$
 is a unit vector pointing toward the sky in the direction of the photons to be observed,
 and
\begin{equation}
 \tilde{f} = \left(
              \begin{array}{c}
               n_\theta \\ n_\phi \\ n_U
              \end{array}
             \right)
             =
              f_0 \left[
                  \left(
                  \begin{array}{c}
                   1 \\ 1 \\ 0
                  \end{array}
                  \right)
                   + \tilde{f}_1
                  \right],
\quad
 \tilde{J}
  = \frac{1}{4\pi} \int_{-1}^1 d\mu' \int_0^{2\pi} d\phi'
    \tilde{P}(\mu,\phi,\mu',\phi') \tilde{f}(\mu',\phi')
\label{def-fluctuation}
\end{equation}
 with $\mu \equiv \cos\theta$.
The components of $\tilde{f}$, $n_\theta$, $n_\phi$ and $n_U$,
 are defined by the squared amplitudes of the electric fields of radiation,
 $I_\theta = E_\theta^2$, $I_\phi = E_\phi^2$ and $U_I = 2 E_\theta E_\phi$,
 divided by $\Delta\nu \cdot h\nu^3/c^2$.
The frequency of CMB radiation
 that we are supposed to observe in this work is $\nu \sim \nu + \Delta\nu$
 with small $\Delta\nu/\nu$.
Note that $\tilde{f}$ is not a vector in space, but simply an array.
The first term of $\tilde{f}$
 is the background contribution of the occupation number of black-body radiation
\begin{equation}
 f_0 = \frac{e^{-h\nu/k_BT_0}}{1-e^{-h\nu/k_BT_0}}
\end{equation}
 and $\tilde{f}_1$ describes perturbations or fluctuations around it.
The right-hand-side of eq.(\ref{eq-rad-transfer}) describes the effect of Thomson scattering with
\begin{equation}
 \tilde{P} = \frac{3}{4}
 \left(
  \begin{array}{ccc}
   \mu^2 \mu'^2 \cos 2 (\phi'-\phi)  & -\mu^2 \cos 2 (\phi'-\phi) & \mu^2 \mu' \sin 2 (\phi'-\phi) \\
   -\mu'^2 \cos 2 (\phi'-\phi)       & \cos 2 (\phi'-\phi)        & - \mu' \sin 2 (\phi'-\phi) \\
   -2 \mu \mu'^2 \sin 2 (\phi'-\phi) & 2 \mu \sin 2 (\phi'-\phi) & 2 \mu \mu' \cos 2 (\phi'-\phi)
  \end{array}
 \right) \ ,
\end{equation}
 and $g(\eta)=\sigma_T n_e(\eta) a(\eta)$,
 where $\sigma_T= (8\pi/3)(\alpha/m_e)^2$ is the Thomson scattering cross section,
 $n_e(\eta)$ is the number density of free electrons,
 and $a(\eta)$ is the scale factor.

The key observation in \cite{Basko:2080} is that for a plane wave perturbation
\begin{equation}
 h_{ij} = D_k(\eta) e^+_{ij} e^{-i k z}
\end{equation}
 with ``plus'' polarization $e^+_{ij} = {\rm diag} (1,-1,0)$, for example,
 the fluctuation $\tilde{f}_1$ is described as
\begin{equation}
 \tilde{f}_{1k} = \alpha_k(\eta,\mu) \, \tilde{a} + \beta_k(\eta,\mu) \, \tilde{b}
\end{equation}
 using specific basis arrays of
\begin{equation}
 \tilde{a} = \frac{1}{2} (1-\mu^2) \cos 2 \phi
             \left(
              \begin{array}{c}
               1 \\ 1 \\ 0
              \end{array}
             \right),
\qquad
 \tilde{b} = \frac{1}{2}
             \left(
              \begin{array}{c}
               (1+\mu^2) \cos 2 \phi \\ -(1+\mu^2) \cos 2 \phi \\ 4\mu\sin 2 \phi
              \end{array}
             \right)
\end{equation}
 by virtue of a special structure of the matrix $\tilde{P}$
 and $e^+_{ij} \hat{n}_i \hat{n}_j = \sin^2\theta \cos 2\phi$.
Therefore,
 the Boltzmann equation reduces to the following two equations
 for $\beta_k$ and $\xi_k \equiv \beta_k + \alpha_k$.
\begin{equation}
 \frac{d\beta_k(\mu)}{d\eta} + ( ik\mu + g(\eta) ) \beta_k(\mu)
 = g(\eta) G_k,
\label{eq-beta}
\end{equation}
\begin{equation}
 \frac{d\xi_k(\mu)}{d\eta} + ( ik\mu + g(\eta) ) \xi_k(\mu)
 = \gamma \frac{d D_k}{d\eta},
\label{eq-xi}
\end{equation}
 where
\begin{equation}
 G_k \equiv
  \frac{3}{16} \int_{-1}^1 d\mu'
  \left[ (1+\mu'^2)^2 \beta_k(\mu') - \frac{1}{2} (1-\mu'^2)^2 \xi_k(\mu') \right]
\label{eq-source}
\end{equation}
 is called the ``source function'', and
\begin{equation}
 \gamma \equiv \frac{\nu}{f_0} \frac{\partial f_0}{\partial\nu}
  = - \frac{h\nu}{k_BT_0}\frac{1}{1-e^{-h\nu/k_BT_0}}\ .
\label{eq-gamma}
\end{equation}
The time dependence of tensor perturbations is determined by
\begin{equation}
 \frac{d^2D_k}{d\eta^2} + \frac{2}{a}\frac{da}{d\eta}\frac{dD_k}{d\eta} + k^2 D_k = 0\ ,
\qquad
 \frac{1}{a^2}\frac{da}{d\eta} = \sqrt{ \frac{\rho}{3 M_P^2} }\ ,
\label{eq-tensor-pert}
\end{equation}
 with the initial conditions given by primordial tensor power spectra
 (neglecting the dumping effect through anisotropic inertia \cite{Weinberg:2003ur}, for simplicity).

Almost the same is true for the case of plane wave perturbation with ``cross'' polarization.
Note that more precisely the tensor perturbations are described as
\begin{equation}
 h_{ij}({\bf x}) = \int \frac{d^3 k}{(2\pi)^3}
  \sum_{\lambda=+, \times}
  \left\{
   \hat{\alpha}_{\bf k} D_k^\lambda e^\lambda_{ij} e^{-i\bf{k}\cdot\bf{x}}
   + \hat{\alpha}_{\bf k}^\dag (D_k^{\lambda})^* (e^\lambda_{ij})^* e^{i\bf{k}\cdot\bf{x}}
  \right\} \ ,
\end{equation}
 with stochastic variables $\hat{\alpha}_{\bf k}$ following
 $\langle \hat{\alpha}_{\bf k} \hat{\alpha}^\dag_{\bf k'} \rangle
  = (2\pi)^3 \delta^3({\bf k} - {\bf k'})$
 and the above $D_k$ should be identified as $D_k^+$.
Every quantity with the dependence of wave number $k$
 is considered to be introduced through the expansion
 with the same stochastic variables $\hat\alpha_{\bf k}$.

The polarization tensor is described as
\begin{equation}
 P_{ab}(\hat{n},k) \equiv \frac{1}{2}
   \left(
   \begin{array}{cc}
    Q_k(\hat{n}) & - U_k(\hat{n}) \sin\theta \\
    -U_k(\hat{n}) \sin\theta & -Q_k(\hat{n}) \sin^2\theta
   \end{array}
   \right),
\end{equation}
 where $Q_k(\hat{n})$ and $U_k(\hat{n})$ are perturbations of Stokes parameters given by
\begin{equation}
 Q_k(\hat{n}) = \frac{T_0}{4} (1 + \mu^2) \cos2\phi \, \beta_k(\mu),
\qquad
 U_k(\hat{n}) = \frac{T_0}{4} 2\mu \sin2\phi \, \beta_k(\mu)
\end{equation}
 for ``plus''--polarized tensor perturbations.
The B-mode portion of the spherical harmonic expansion of the polarization tensor is
\begin{equation}
 a^{\rm B}_{\ell m}(k) =
  N_\ell \int d{\hat n} \, \nabla_a \nabla_c \, \epsilon^c{}_b \, P^{ab}(\hat{n},k)
   \left( Y^m_\ell(\hat{n}) \right)^*
\end{equation}
 with $N_\ell = \sqrt{2(\ell-2)!/(\ell+2)!}$
 and metric $g_{ab} = {\rm diag} ( 1, \sin^2\theta )$,
 and then
\begin{equation}
 C^{\rm BB}_\ell
  = 2 \int \frac{d^3k}{(2\pi)^3} \frac{1}{2\ell+1} \sum_m | a_{\ell m}^{\rm B}(k) |^2
  = \frac{T_0^2}{4\pi}
    \int dk k^2
    \left\vert
     \frac{\ell+2}{2\ell+1} \beta_{k, \ell-1}
     +
     \frac{\ell-1}{2\ell+1} \beta_{k, \ell+1}
    \right\vert^2,
\label{eq-C^BB}
\end{equation}
 where
\begin{equation}
 \beta_k(\mu) = \sum_\ell (2\ell + 1) \beta_{k, \ell} P_\ell(\mu)\ ,
\qquad
 \xi_k(\mu) = \sum_\ell (2\ell + 1) \xi_{k, \ell} P_\ell(\mu)
\end{equation}
 are expansion in Legendre polynomials.
In eq.(\ref{eq-C^BB}) we have assumed a homogeneous and isotropic Universe,
 and the two polarizations of tensor perturbations contribute the same amount (parity conservation).

The magnitude of tensor perturbation $D_k$ determines the magnitude of $\xi_k$ through eq.(\ref{eq-xi}).
That $\xi_k$ determines in part the magnitude of the source function $G_k$ of eq.(\ref{eq-source}),
 and eq.(\ref{eq-beta}) determines the magnitude of $\beta_k$.
This is how the magnitude of $C^{\rm BB}_\ell$ is determined.
The polarization of perturbations, $\beta_k$, should vanish,
 if tensor perturbations vanish, or with vanishing $g$,
 in the absence of Thomson scattering with free electrons.
This fact is clearly seen
 in the formal solutions of eqs.(\ref{eq-beta}) and (\ref{eq-xi})
\cite{Zaldarriaga:1996xe},
\begin{equation}
 \beta_k(\eta,\mu)
  = \int_0^\eta d\eta' e^{-\kappa(\eta,\eta')} e^{-ik\mu(\eta-\eta')} g(\eta') G_k(\eta')\ ,
\label{eq-formal-sol-beta}
\end{equation}
\begin{equation}
 \xi_k(\eta,\mu)
  = \int_0^\eta d\eta' e^{-\kappa(\eta,\eta')} e^{-ik\mu(\eta-\eta')} \gamma \dot{D}_k(\eta')\ ,
\end{equation}
 where $\kappa(\eta,\eta')$ is defined as
\begin{equation}
 \frac{d}{d\eta}\kappa(\eta,\eta') = g(\eta)
\quad \mbox{and} \quad
 \kappa(\eta,\eta) = 0\ ,
\end{equation}
 and the initial conditions are $\beta_k(0)=0$ and $\xi_k(0)=0$.
An additional approximation identifies $\eta=0$ with the time at the end of inflation
 and also at the beginning of recombination.
The equations for the expansion coefficients in Legendre polynomials are
\begin{equation}
 \dot{\beta}_{k,0} = - g \beta_{k,0} - ik \beta_{k,1} + g G_k\, ,
\qquad
 \dot{\beta}_{k,\ell}
  = - g \beta_{k,\ell} - \frac{ik}{2\ell+1} [ \ell \beta_{k,\ell-1} + (\ell+1) \beta_{k,\ell+1} ]\, ,
\label{eq-beta-ell}
\end{equation}
\begin{equation}
 \dot{\xi}_{k,0} = - g \xi_{k,0} - ik \xi_{k,1} + \gamma \dot{D}_k\, ,
\qquad
 \dot{\xi}_{k,\ell}
  = - g \xi_{k,\ell} - \frac{ik}{2\ell+1} [ \ell \xi_{k,\ell-1} + (\ell+1) \xi_{k,\ell+1} ]\, ,
\label{eq-xi-ell}
\end{equation}
 where $\ell>0$, dots indicate derivatives with respect to $\eta$, and
\begin{equation}
 G_k = \frac{7}{10} \beta_{k,0} + \frac{5}{7} \beta_{k,2} + \frac{3}{35} \beta_{k,4}
 - \frac{1}{10} \xi_{k,0} + \frac{1}{7} \xi_{k,2} - \frac{3}{70} \xi_{k,4}\, .
\end{equation}
The formal solution eq.(\ref{eq-formal-sol-beta}) gives
\begin{equation}
 \beta_{k,\ell}
  = \int_0^\eta d\eta' e^{-\kappa(\eta,\eta')} \, i^\ell j_\ell(k(\eta'-\eta)) \, g(\eta') G_k(\eta')\, ,
\label{eq-formal-sol-beta-ell}
\end{equation}
 where $j_\ell$ is a spherical Bessel function.

How can one approximate the source function $G_k$
 to investigate $C^{\rm BB}_\ell$, which is described by $\beta_{k,\ell}$
 as in eq.(\ref{eq-C^BB})?
The following truncation in eqs.(\ref{eq-beta-ell}) and (\ref{eq-xi-ell}),
 which can be a good approximation for $g \gg k$
 (the tight coupling limit \cite{Zaldarriaga:1995gi}), is a method to obtain $G_k$.
\begin{equation}
 \dot{\beta}_{k,0} \simeq - g \beta_{k,0} + g G_k\, ,
\qquad
 \dot{\beta}_{k,\ell} \simeq - g \beta_{k,\ell}\,,
\end{equation}
\begin{equation}
 \dot{\xi}_{k,0} \simeq - g \xi_{k,0} + \gamma \dot{D}_k\,,
\qquad
 \dot{\xi}_{k,\ell} \simeq - g \xi_{k,\ell}\,.
\end{equation}
One can reasonably expect that $\beta_\ell$ and $\xi_\ell$ for $\ell>0$
 be exponentially smaller than $\beta_0$ and $\xi_0$, respectively.
Setting $\beta_\ell=0$ and $\xi_\ell=0$ for $\ell>0$, the equations become
\begin{equation}
 \dot{\beta}_{k,0} \simeq - \frac{3}{10} g \beta_{k,0} - \frac{1}{10} g \xi_{k,0}\,,
\qquad
 \dot{\xi}_{k,0} \simeq - g \xi_{k,0} + \gamma \dot{D}_k
\end{equation}
 with the source function $G_k = (7/10) \beta_{k,0} - (1/10) \xi_{k,0}$.
These equations determine a differential equation for the source function
\begin{equation}
 \dot{G}_k \simeq - \frac{3}{10} g G_k - \frac{1}{10} \gamma \dot{D}_k\,,
\end{equation}
 whose formal solution is
\begin{equation}
 G_k(\eta) \simeq
  - \frac{1}{10} \int_0^\eta d\eta' e^{-\frac{3}{10} \kappa(\eta,\eta')} \gamma \dot{D}_k(\eta')\, .
\end{equation}

Now we concentrate on the polarization
 induced by the scattering with free electrons produced by reionization.
The function $g(\eta) = \sigma_T n_e(\eta) a(\eta)$ vanishes
 before the time of reionization $\eta_{\rm ion}$, which corresponds to $z_{\rm ion} \simeq 8$.
The source function should also vanish before $\eta_{\rm ion}$.
Therefore,
\begin{equation}
 \beta_{k,\ell}(\eta_0)
  = \int_{\eta_{\rm ion}}^{\eta_0} d\eta'
    e^{-\kappa(\eta_0,\eta')} \, i^\ell j_\ell(k(\eta'-\eta_0)) \, g(\eta') G_k(\eta')\ ,
\label{eq-beta-formal}
\end{equation}
\begin{equation}
 G_k(\eta)
  \simeq - \frac{1}{10} \int_{\eta_{\rm ion}}^{\eta} d\eta'
           e^{-\frac{3}{10} \kappa(\eta,\eta')} \gamma \dot{D}_k(\eta').
\label{eq-source-formal}
\end{equation}
Furthermore, if the functions
 $\exp(-\kappa(\eta_0,\eta'))$ and $\exp(-(3/10)\kappa(\eta,\eta'))$ are slowly varying and almost unity
 (this is the case after reionization, which will be discussed in the next section),
 the preceding equations become
\begin{equation}
 \beta_{k,\ell}(\eta_0)
  \simeq \int_{\eta_{\rm ion}}^{\eta_0} d\eta'
         i^\ell j_\ell(k(\eta'-\eta_0)) \, g(\eta') G_k(\eta')\ ,
\end{equation}
\begin{equation}
 G_k(\eta)
  \simeq - \frac{1}{10} \gamma \left( D_k(\eta) - D_k({\eta_{\rm ion}}) \right)\ ,
\end{equation}
 and one is thus led to a simple result:
\begin{equation}
 \beta_{k,\ell}(\eta_0)
  \simeq i^\ell (-1)^{\ell+1} \frac{1}{10} \gamma
         \int_{\eta_{\rm ion}}^{\eta_0} d\eta'
         \, j_\ell(k(\eta_0-\eta')) \, g(\eta')
         \left( D_k(\eta') - D_k({\eta_{\rm ion}}) \right)\ .
\label{eq-beta-master}
\end{equation}
The polarization of perturbations, $\beta_{k,\ell}$,
 follows directly from the scattering rate $g(\eta)$ and the tensor perturbations $D_k(\eta)$.

\begin{figure}[t]
\centering
\includegraphics[width=60mm]{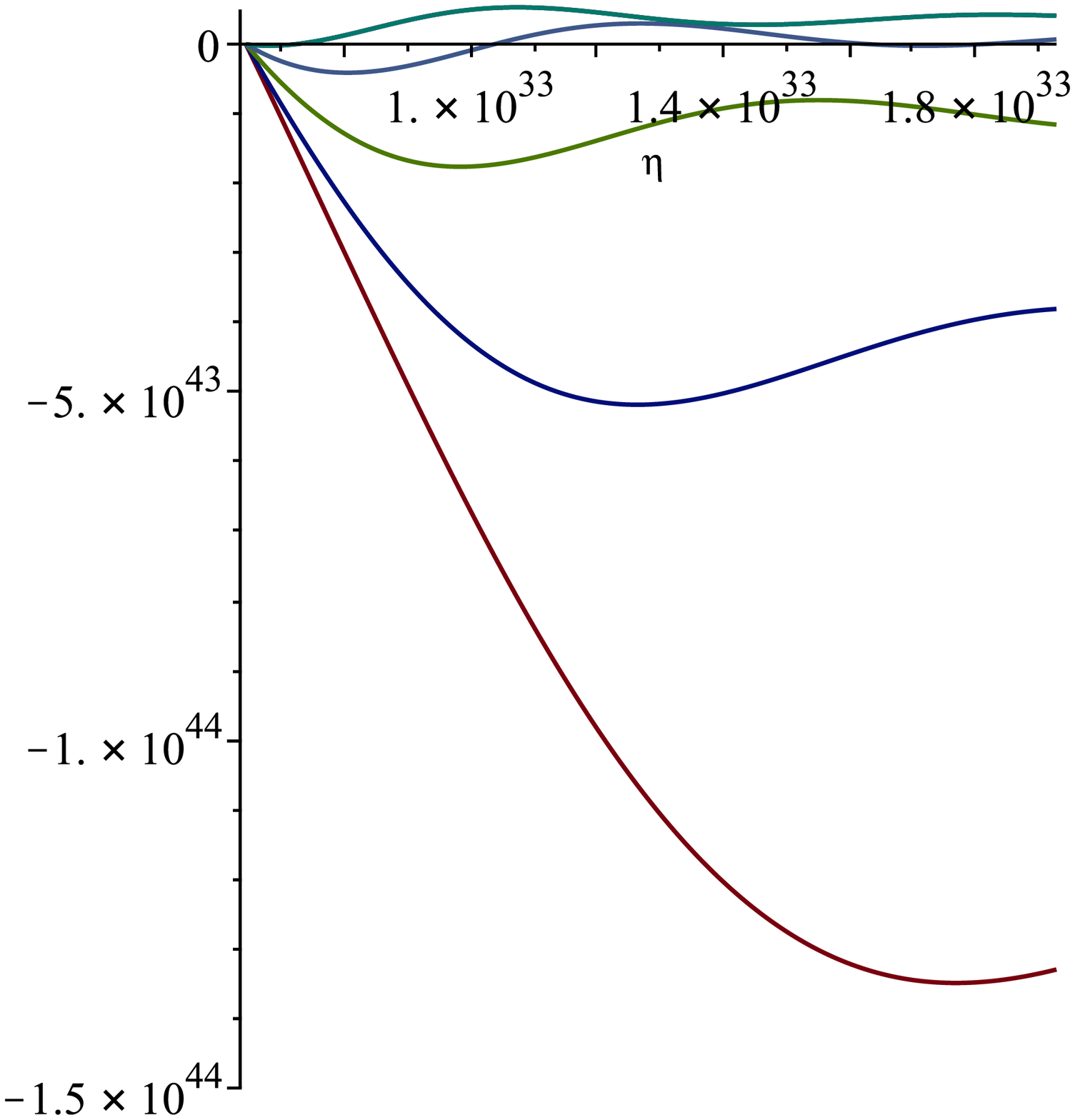}
\qquad
\includegraphics[width=60mm]{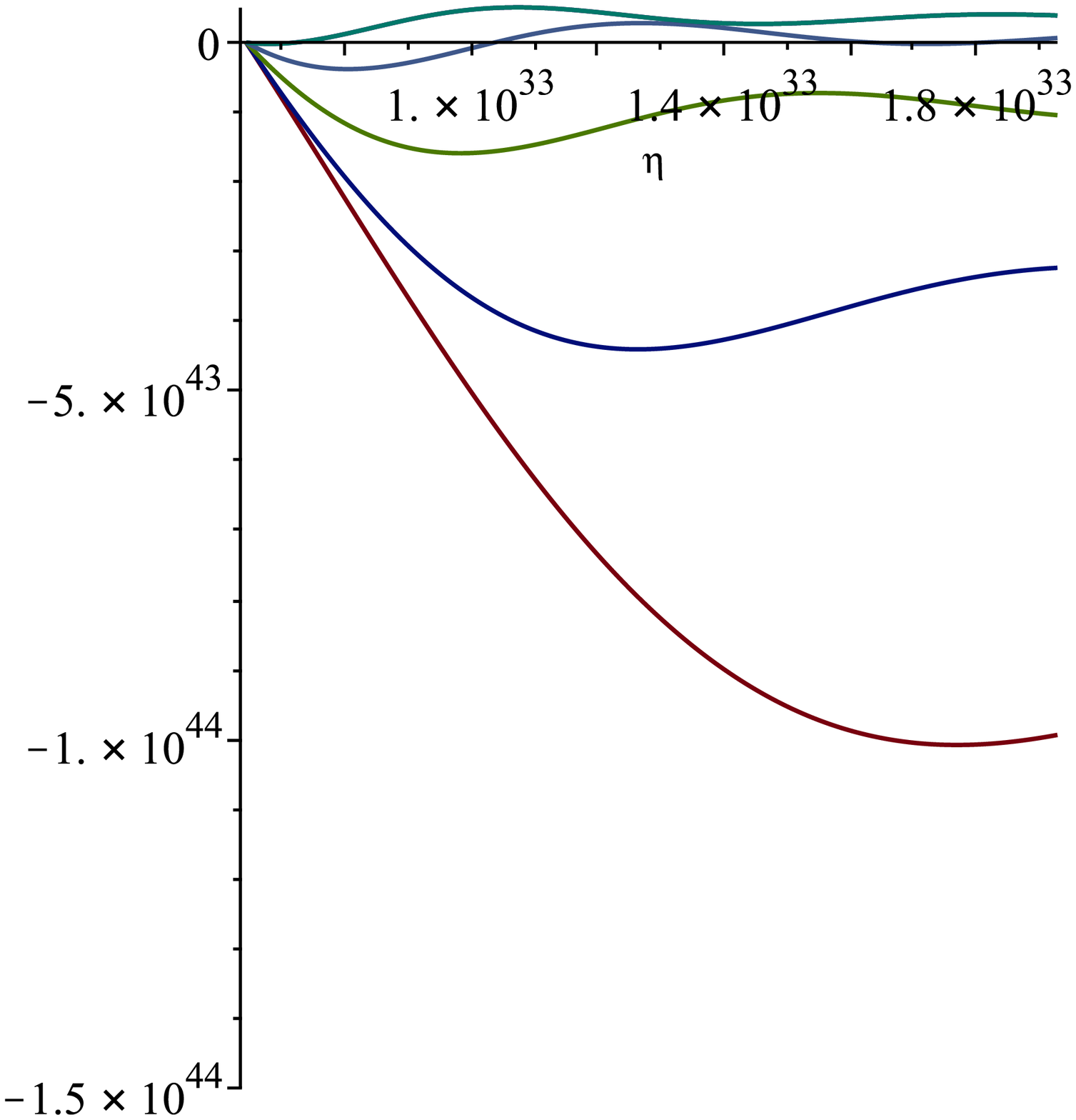}
\caption{
Left:
$D_k(\eta)-D_k(\eta_{\rm ion})$ [$\rm{eV}^{-3/2}$]
 as functions of $\eta$ [$\rm{eV}^{-1}$]
 for various values of $k_{\rm min}=2\pi/\eta_0$ to $k_{\rm max}=2\pi/\eta_{\rm ion}$
 (from lowest line to upper lines) with $\Delta=0$ ($\Lambda$CDM model).
Right:
$D_k(\eta)-D_k(\eta_{\rm ion})$  [$\rm{eV}^{-3/2}$]
 as functions of $\eta$ [$\rm{eV}^{-1}$]
 for various values of $k_{\rm min}=2\pi/\eta_0$ to $k_{\rm max}=2\pi/\eta_{\rm ion}$
 (from lowest line to upper lines) with $\Delta=0.351 \times 10^{-3}$ [$\rm{Mpc}^{-1}$].
}
\label{fig:tensor-pert}
\end{figure}

The tensor perturbations $D_k(\eta)$ are obtained
 solving the first equation in eq.(\ref{eq-tensor-pert})
 taking into account the background expansion of the Universe,
 which is determined by the second equation in eq.(\ref{eq-tensor-pert}),
 and with the initial condition determined by the primordial tensor power spectrum.
Since the Universe is matter dominated, $\rho \propto 1/a^3$, in the reionization era
 one can approximate the scale factor as $a(t) = (t/t_0)^{2/3}$,
 where $t_0$ is the age of the Universe.
The relation between coordinate time $t$ and conformal time $\eta$ is then
 $\eta/\eta_0 = (t/t_0)^{1/3}$, and $a(\eta) = (\eta/\eta_0)^2$ with $\eta_0 = 3 t_0$.
The beginning of reionization
 is determined making use of the relation $1+z=1/a$ with $z_{\rm ion}=8$
 as $\eta_{\rm ion} = \eta_0/3$.
The initial condition is set at $\eta=0$,
 approximately the time at the end of inflation, as
\begin{equation}
 D_k(0) = \sqrt{\frac{2 \pi^2 A_T}{k^3}}
\qquad \mbox{and} \qquad
 \dot{D}_k(0) = 0
\end{equation}
 for the $\Lambda$CDM model,
 assuming for the spectral index of primordial tensor power spectrum $n_T=0$
 (no $k$ dependence, which is predicted by typical inflation models),
 where $A_T = r_{TS} A_S$ with tensor-to-scalar ratio $r_{TS}$
 and the amplitude scalar perturbation $A_S \simeq 2.1 \times 10^{-9}$ \cite{Aghanim:2018eyx}.
We can introduce the $\Delta$ parameter of eq.(\ref{primordial-tensor-Delta}) taking
\begin{equation}
 D_k(0) = \sqrt{\frac{2 \pi^2 A_T}{(k^2+\Delta^2)^{3/2}}} \ ,
\end{equation}
 and other possible primordial power spectra can be treated in the same way.
We approximately neglect
 the radiation dominated era and the recent accelerated expansion of the Universe, for simplicity.
In this setting the solution of eq.(\ref{eq-tensor-pert}) is then
\begin{equation}
 D_k(\eta)
  = \sqrt{\frac{2 \pi^2 A_T}{(k^2+\Delta^2)^{3/2}}} \cdot 3 \sqrt{\frac{\pi}{2}}
    (k\eta)^{-3/2} J_{3/2}(k\eta)\ .
\end{equation}
We may set $\Delta=0$ for the $\Lambda$CDM model.
Fig.\ref{fig:tensor-pert} shows the resulting behaviors of $D_k(\eta)-D_k(\eta_{\rm ion})$
 that appear in eq.(\ref{eq-beta-master}),
 as functions of $\eta$ for relevant values of $k$.
The magnitudes of perturbations with smaller wave numbers (and thus longer wave lengths or large scales)
 are larger than that with larger wave numbers.
Tensor perturbations with smaller wave numbers
 are diluted by the expansion of the Universe for a shorter time than those with larger wave numbers,
 and the initial magnitudes of tensor perturbations with smaller wave numbers
 are larger than those with larger wave numbers.
The magnitudes of large--scale perturbations are suppressed by the effect of non-zero $\Delta$.

\section{The Results and their Indications}
\label{sec:results}

The last step in the calculation with eq.(\ref{eq-beta-master})
 is to specify the function $g(\eta) = \sigma_T n_e(\eta) a(\eta)$,
 especially the time evolution of the free electron density $n_e(\eta)$ in the process of reionization.
We do not consider the free electrons that remain after recombination,
 since they have been diluted by the expansion of the Universe
 and do not dominantly contribute to $C^{BB}_\ell$ for $\ell \lesssim 10$.
Since the reionization process is not known in detail \cite{Adam:2016hgk},
 we are going to explore two extreme cases:
 instantaneous reionization with $n_e a^3$ a step function
 and slow reionization with $n_e a^3$ increasing slowly
 (note that free electron density is diluted as $n_e \sim 1/a^3$ by the expansion of the Universe).
An important constraint is that
\begin{equation}
 \tau \simeq \int_{\eta_{\rm ion}}^{\eta_0} d\eta \, g(\eta)
\label{eq-optical-depth}
\end{equation}
  is the optical depth, which is determined as $\tau = 0.054 \pm 0.007$ \cite{Aghanim:2018eyx}.
Furthermore,
 the reionization process should terminate by the time corresponding to $z \simeq 6$, consistently with
 the observation of Lyman-alpha forest and Gunn-Peterson trough \cite{Gunn:1965hd}.
\begin{figure}[t]
\centering
\includegraphics[width=50mm]{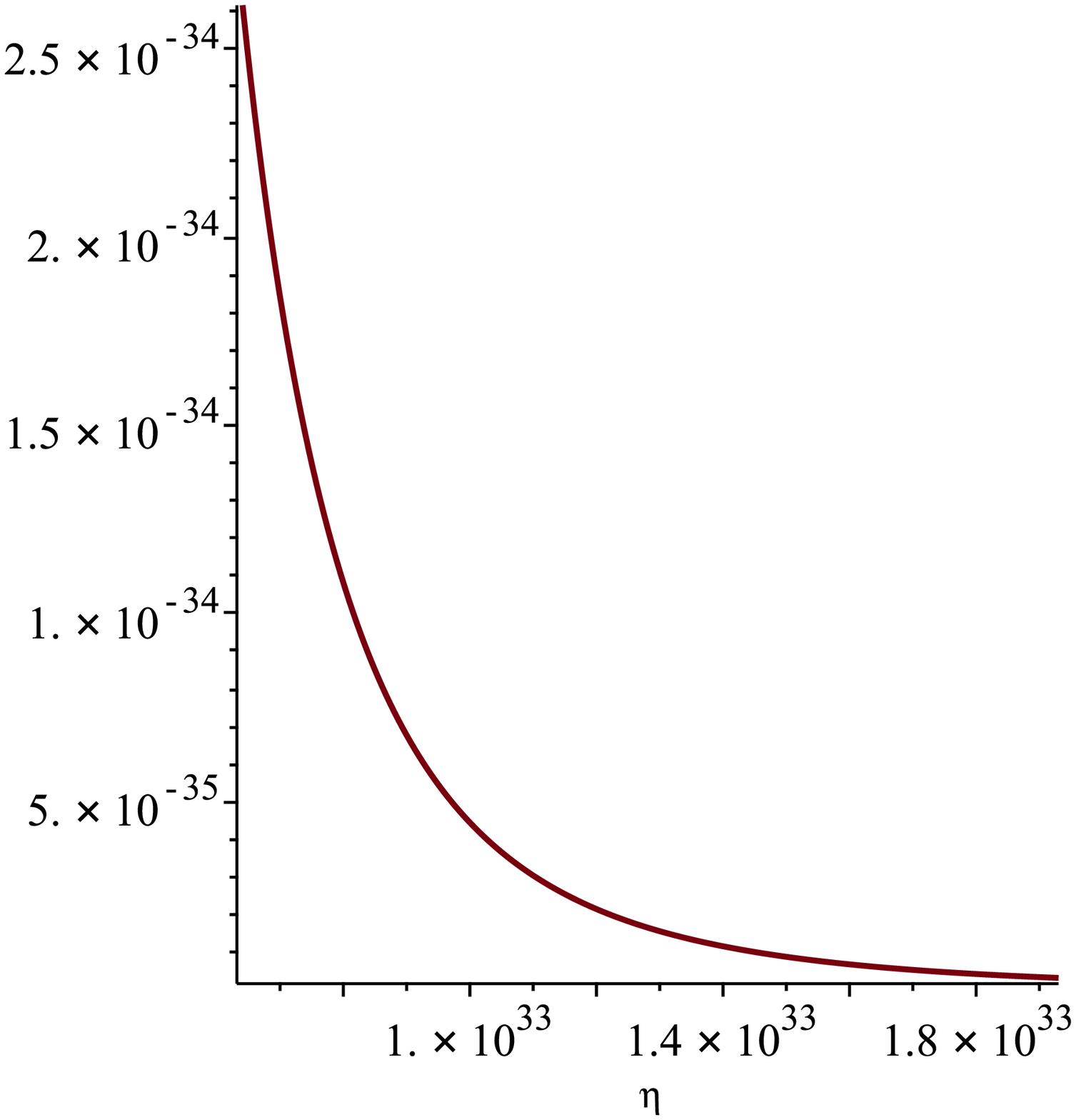}
\quad
\includegraphics[width=50mm]{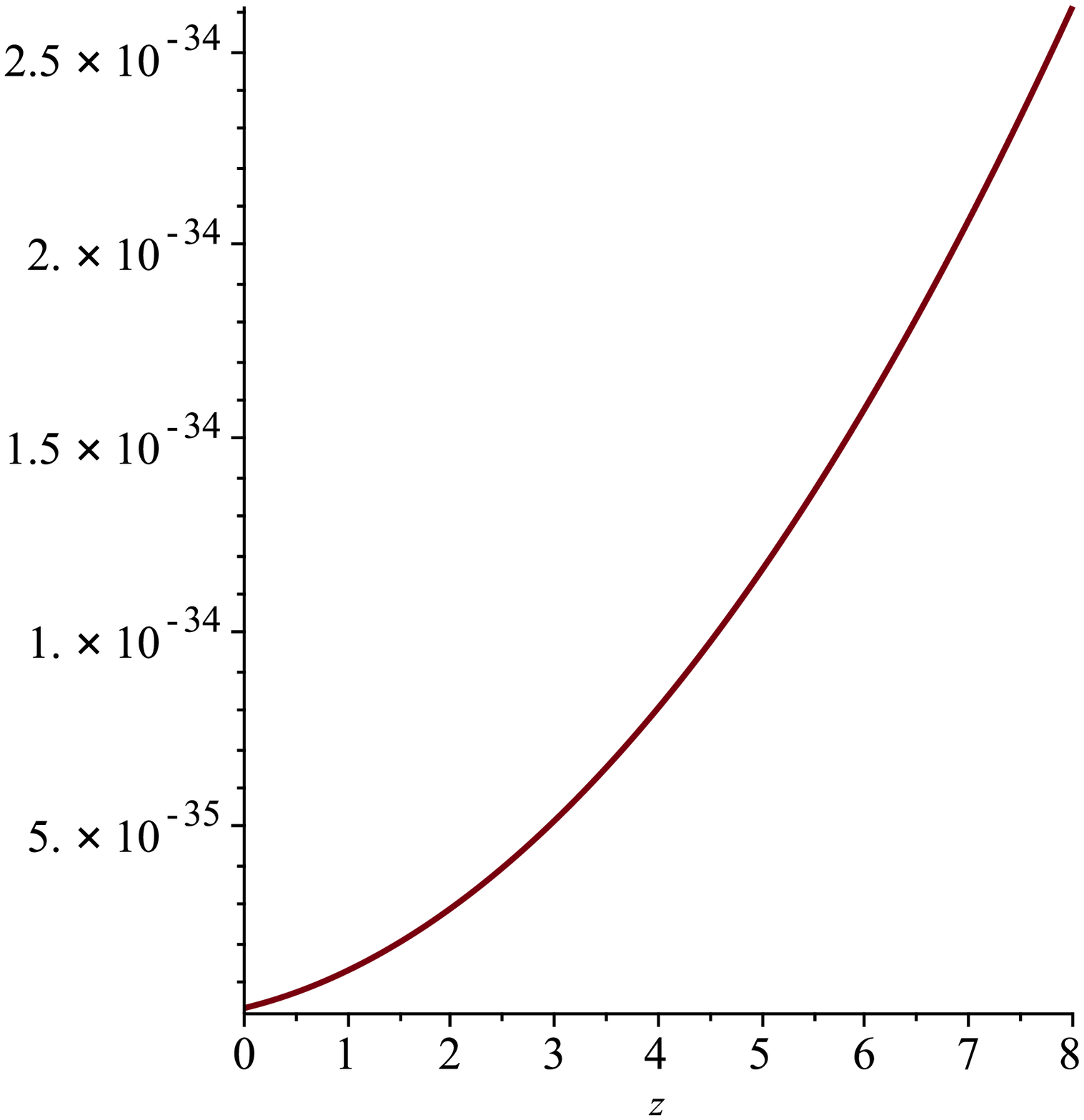}
\qquad
\includegraphics[width=50mm]{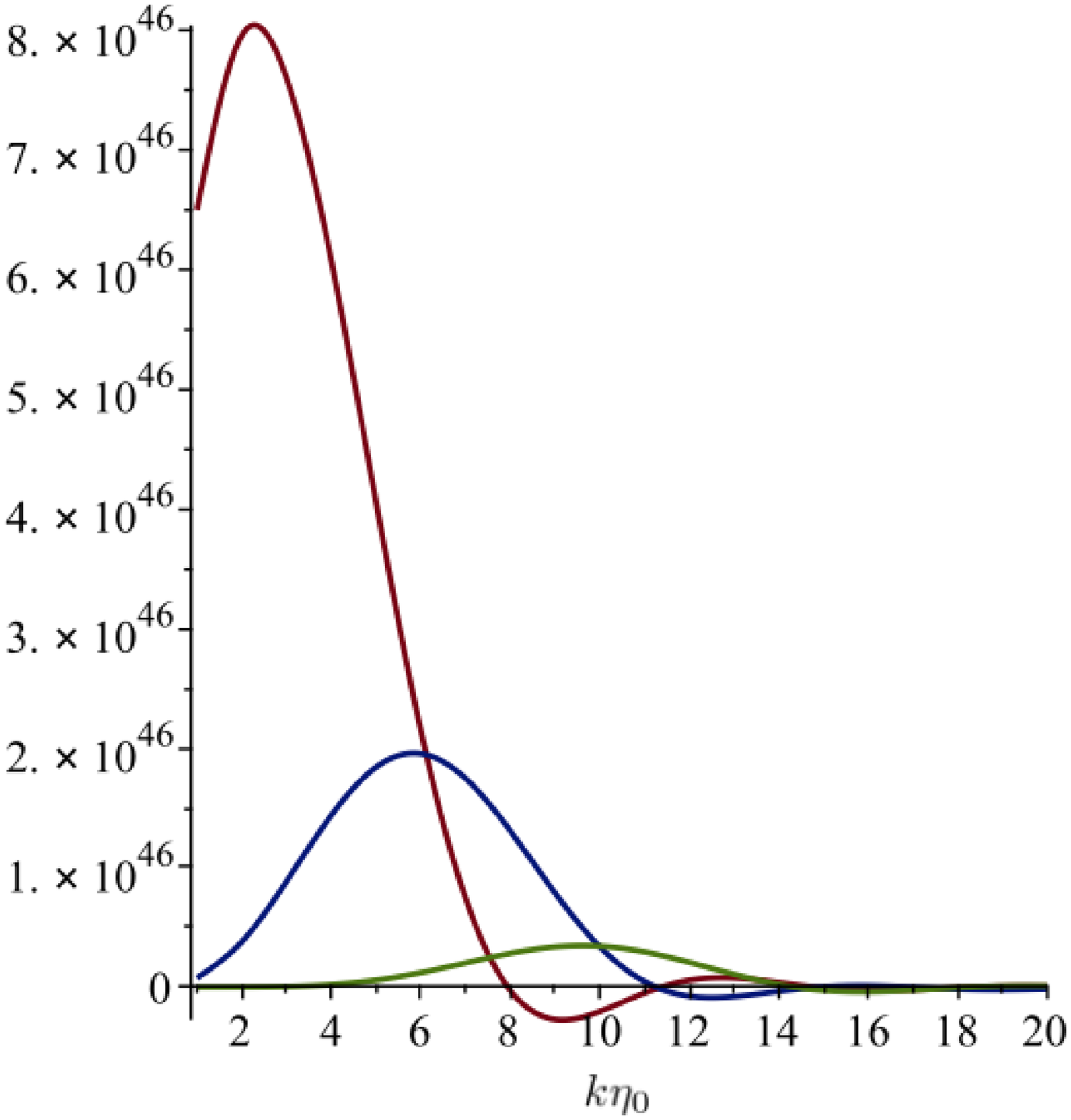}
\caption{
Left:
$g$ [$\rm{eV}$] in case of instantaneous reionization
 as a function of conformal time $\eta$ [$\rm{eV}^{-1}$].
Middle:
 $g$ [$\rm{eV}$] in case of instantaneous reionization
 as a function of redshift $z$.
Right:
$\beta_{k,\ell}/i^\ell(-1)^{\ell+1}$ [$\rm{eV}^{-3/2}$] as functions of $k\eta_0$
 for $\ell=0,2,5$ corresponding to high to low peaks
 (in case of instantaneous reionization with $\Delta=0$ and $\gamma=-2.8$).
}
\label{fig:g-and-beta}
\end{figure}

In case of instantaneous reionization we may set $n_e(\eta) = n^{\rm ion}_e / a(\eta)^3$,
 and $\tau = 0.054$ gives $n_e^{\rm ion} \simeq 1.9 \times 10^{-21}$ [eV${}^3$] with $z_{\rm ion}=8$
 (see left panel of fig.\ref{fig:g-and-beta}).
In this case
\begin{equation}
 \kappa(\eta,\eta')
  = \frac{1}{3} \sigma_T n_e^{\rm ion} \eta_0
    \left( (\eta_0/\eta')^3 - (\eta_0/\eta)^3 \right)\ ,
\end{equation}
 and the numerical value of a factor in eq.(\ref{eq-beta-formal}) is
\begin{equation}
 e^{-\kappa(\eta_0,\eta')} = 1 + {\cal O}(10^{-2})
\end{equation}
 in the relevant range $\eta_{\rm ion} < \eta < \eta_0$,
 so that the approximation in the previous section is justified.
The same is true for the factor $\exp(-(3/10)\kappa(\eta,\eta'))$ in eq.(\ref{eq-source-formal})
 in the relevant range $\eta_{\rm ion} < \eta, \eta' < \eta_0$.
Here, we investigate the typical wave number $k_{\rm ion}$,
 which corresponds to the horizon scale at the time of reionization,
 to understand whether the truncation can be a valid approximation ($g(\eta_{\rm ion}) \gg k_{\rm ion}$).
The comoving size of the particle horizon at $\eta=\eta_{\rm ion}$ is
\begin{equation}
 d_H(\eta_{\rm ion})
  = \int_0^{t_{\rm ion}} \frac{dt}{a(t)} = \int_0^{\eta_{\rm ion}} d\eta = \eta_{\rm ion}.
\end{equation}
The comoving distance from the present observer, which is measured by light propagation, is
\begin{equation}
 r_H(\eta_{\rm ion}) = \int_{\eta_{\rm ion}}^{\eta_0} d\eta = \eta_0 - \eta_{\rm ion} \simeq \eta_0\ .
\end{equation}
Therefore, the resulting angular separation at present is
 $\Delta\theta_{\rm ion} \simeq d_H/r_H \simeq \eta_{\rm ion}/\eta_0$.
Since the corresponding multipole is
 $\ell_{\rm ion} \simeq \pi/\Delta\theta_{\rm ion} \simeq \pi \eta_0/\eta_{\rm ion}$,
 the comoving wave number that corresponds to the scale of horizon
 at the beginning of reionization is
\begin{equation}
 k_{\rm ion} = \frac{2\pi}{d_H(\eta_{\rm ion})} \simeq \frac{2}{\eta_0} \ell_{\rm ion}
 \simeq 1.0 \times 10^{-33} \, \ell_{\rm ion} \, \rm{[ev]}.
\end{equation}
This is about three times larger than $g(\eta_{\rm ion}) \simeq 2.6 \times 10^{-34}$ [eV],
 and the truncation is not a good approximation.
But the truncation is not a terribly bad approximation either,
 because the peak value of $|\beta_\ell|$ is larger than that of $|\beta_{\ell'}|$
 for $\ell < \ell'$ (see the right panel of fig.\ref{fig:g-and-beta}),
 and the same is true for $|\xi_\ell|$.
Therefore,
 the truncation is a worthwhile procedure to provide a simple understanding of the basic Physics,
 although the precision is somewhat sacrificed.

The value of $\gamma$, which is defined in eq.(\ref{eq-gamma}),
 depends on the frequency of the observed CMB radiation\footnote{
Notice that
 the frequency dependence is introduced in the general perturbation in eq.(\ref{def-fluctuation})
  by the setup of \cite{Keating:1997cv}, to which we are referring.
}.
The higher frequencies give larger $|\gamma|$.
The frequencies for polarization measurements by PLANCK
 are $30, 44, 70, 100, 143, 217$ and $353$ [GHz] \cite{Akrami:2018vks},
 and the corresponding frequency range by LiteBIRD, for example,
 will be $40 \sim 400$ [GHz] \cite{Ishino:2016wxl}.
In this paper we take $\nu=150$ [GHz] and $\gamma \simeq -2.8$.

\begin{figure}[t]
\centering
\includegraphics[width=50mm]{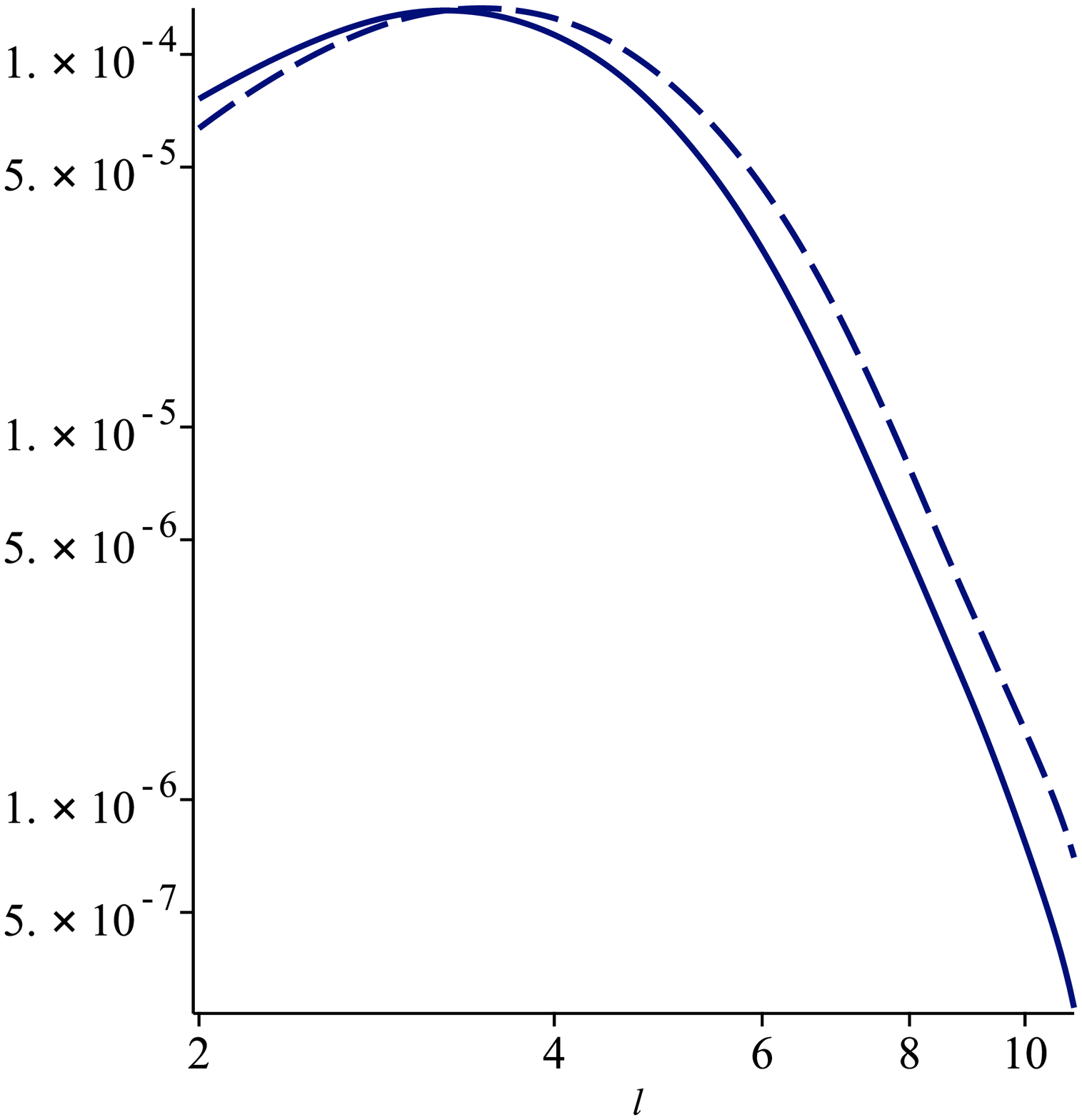}
\quad
\includegraphics[width=50mm]{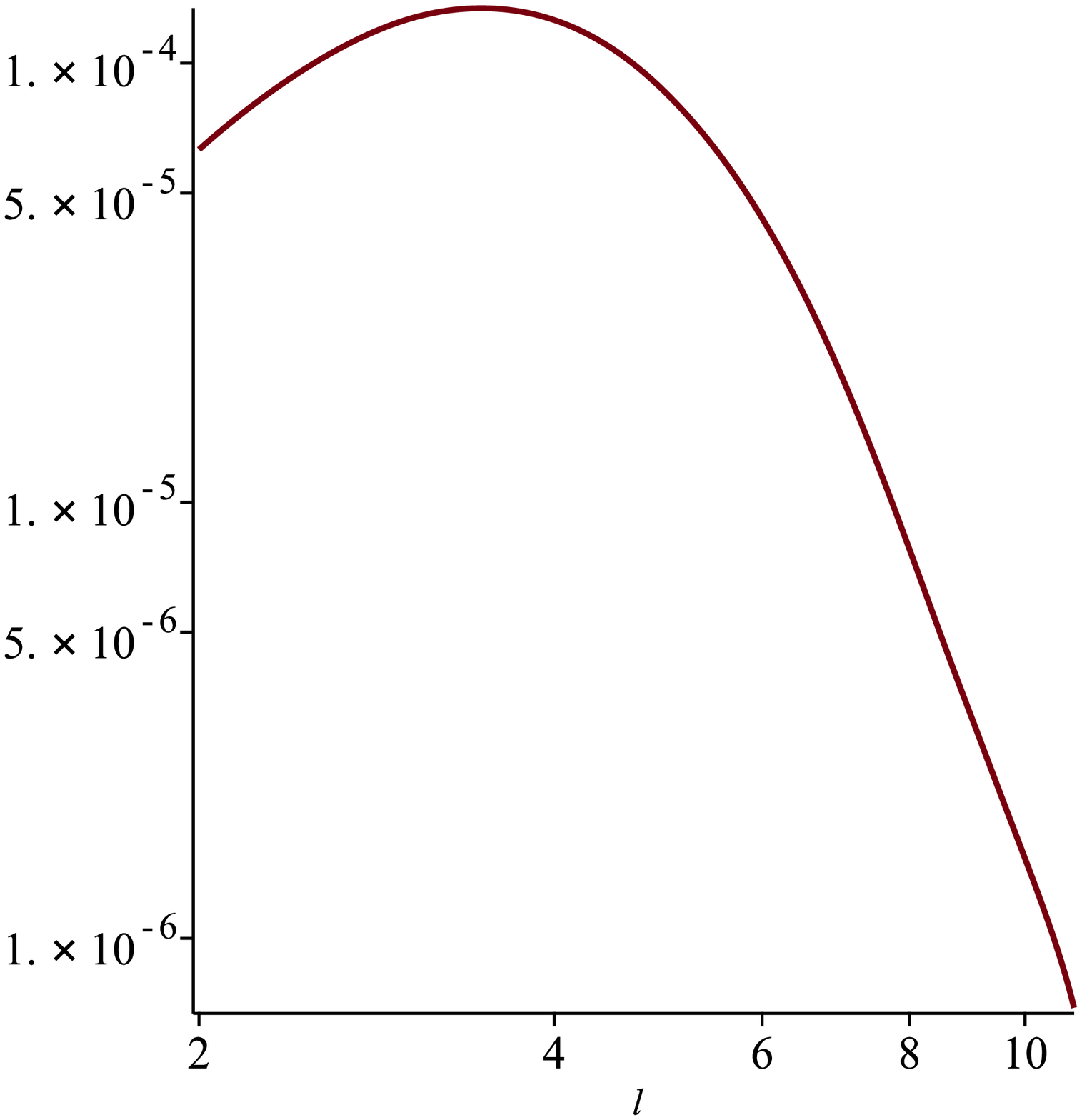}
\quad
\includegraphics[width=50mm]{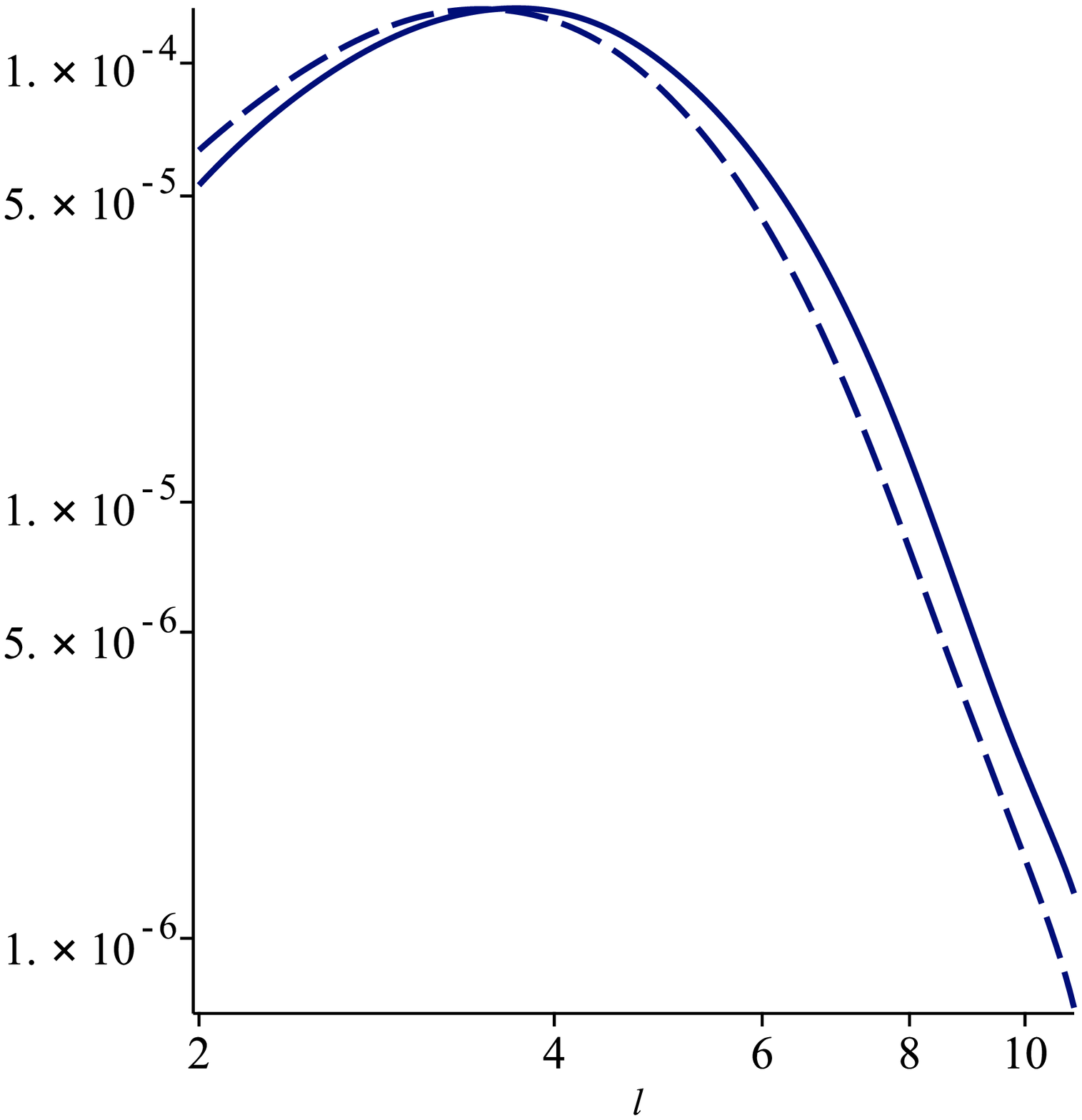}
\quad
\includegraphics[width=50mm]{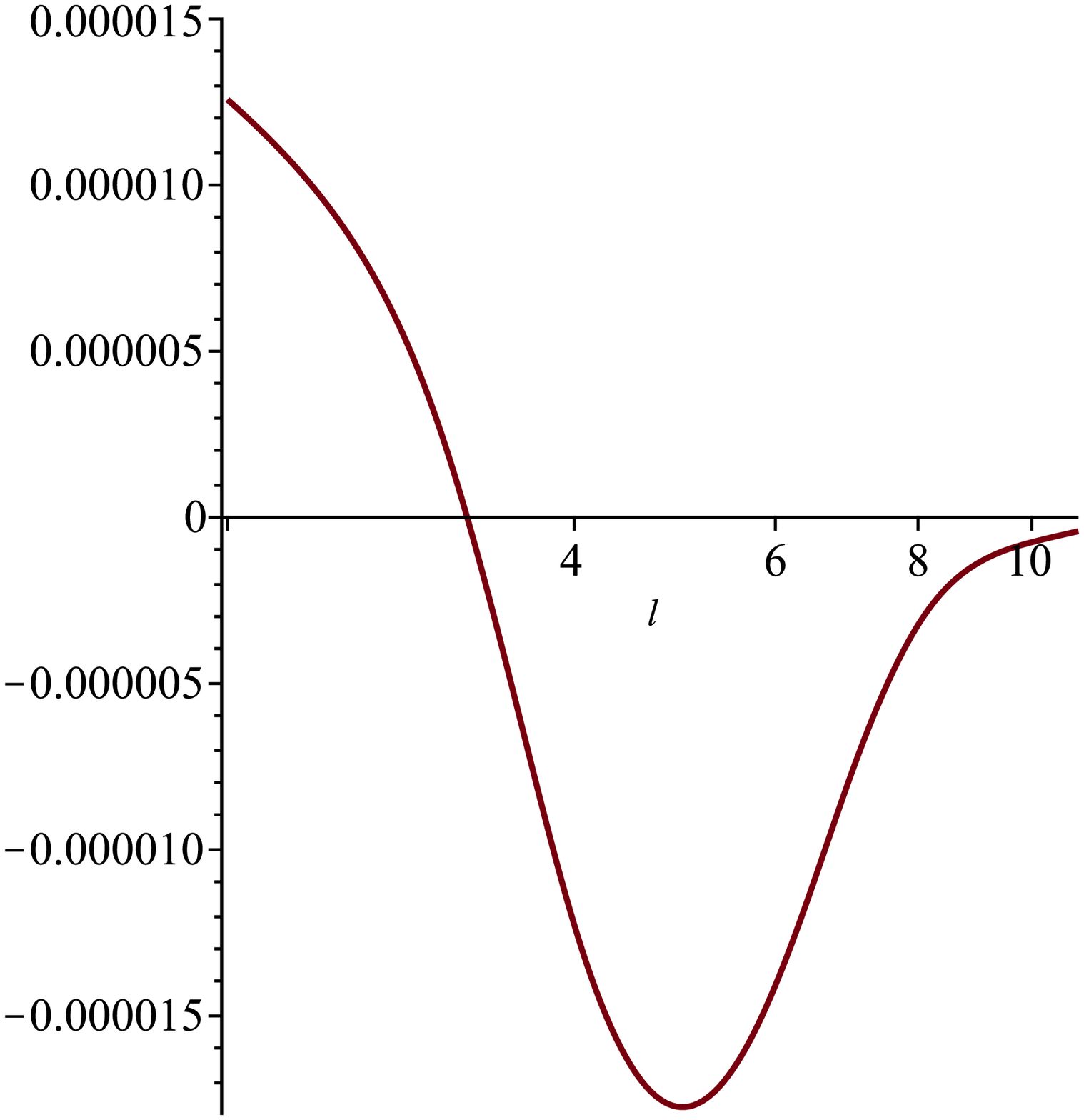}
\quad
\includegraphics[width=50mm]{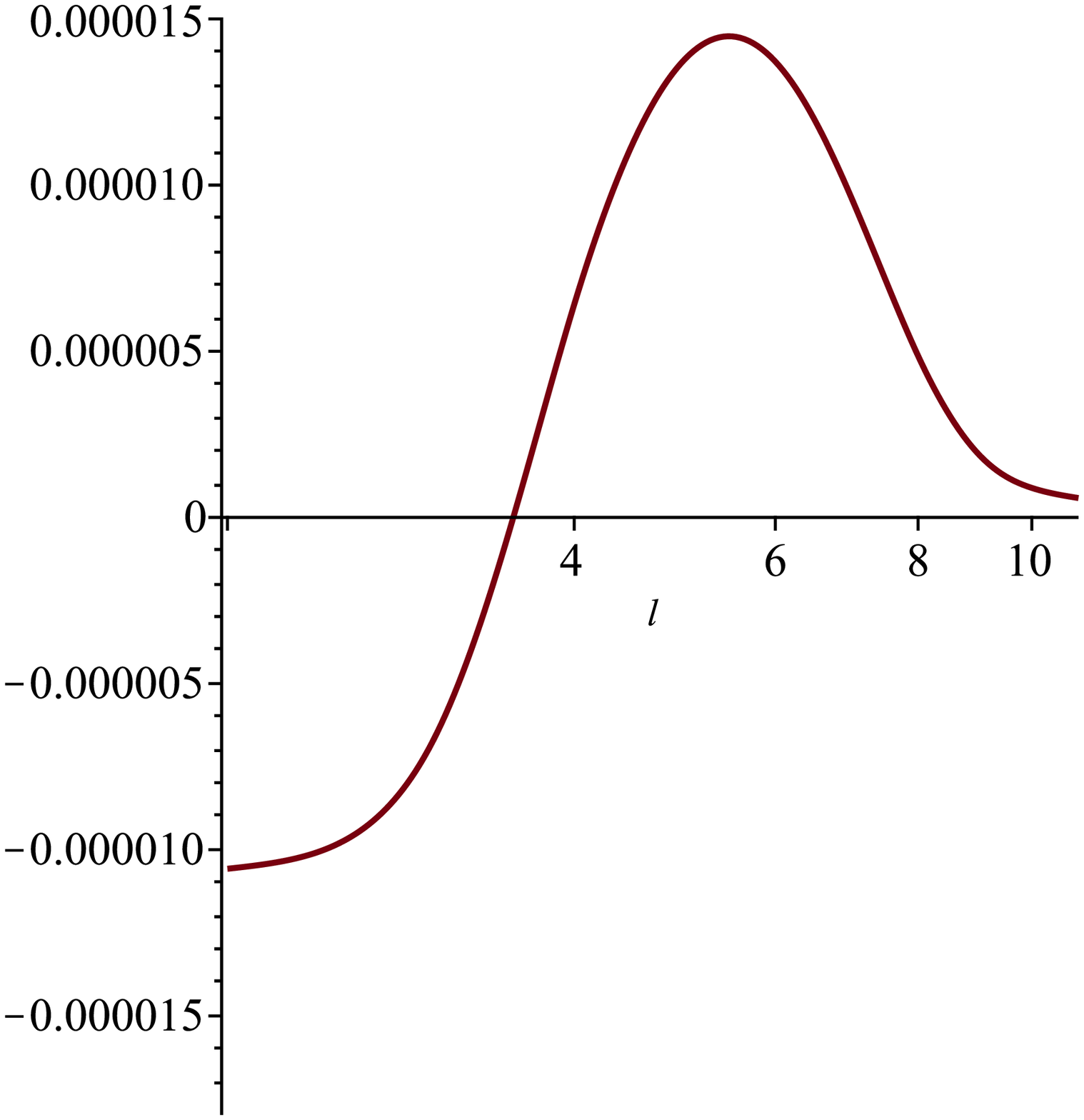}
\caption{
The plots in upper line show
 $D^{\rm BB}_\ell = (\ell(\ell+1)/2\pi) C^{\rm BB}_\ell$ [$\mu {\rm K}^2$] ($\Delta=0$ case)
 under instantaneous reionization with various red shift values at the start of reionization:
 $z_{\rm ion}=7$, $z_{\rm ion}=8$ and $z_{\rm ion}=9$ from left to right, respectively,
 where dashed lines in the left and right panels correspond to $z_{\rm ion}=8$.
The plots in the lower line show
 the difference between the $z_{\rm ion} = 7$ and $z_{\rm ion} = 8$ cases
 and the differences between the $z_{\rm ion}=9$ and $z_{\rm ion}=8$ cases,
 corresponding to the left and right plots in the upper line, respectively.
}
\label{fig:D^BB_ell-inst}
\end{figure}

The result for the B-mode polarization power spectrum $D^{\rm BB}_\ell$
 for $\Lambda$CDM model is displayed in the middle panel of fig.\ref{fig:D^BB_ell-inst}.
In this paper we take $r_{TS}=0.03$ and $A_S = 2.1 \times 10^{-9}$ for all numerical calculations.
The result is consistent with more precise numerical calculations,
 even in a certain quantitative level (see \cite{Ishino:2016wxl}, for example).
It can be simply understood looking at eq.(\ref{eq-beta-master}) (and eq.(\ref{eq-C^BB}))
while taking into account the well-known behavior of spherical Bessel functions $j_\ell(x)$
 that choose smaller $x$ for smaller $\ell$ and vice versa.
For larger $\ell$, which correspond to smaller scales or larger wave numbers,
 the magnitude of the resulting power is small because of the smaller magnitude of tensor perturbations
 (see fig.\ref{fig:tensor-pert}).
For smaller values of $\ell$, which correspond to larger scales or smaller wave numbers,
 the power is also small because of smaller free electron density
 (see left panel of fig.\ref{fig:g-and-beta}).
Therefore, a peak is present for a certain value of $\ell$.

The left and right panels of fig.\ref{fig:D^BB_ell-inst}
 display $D^{\rm BB}_\ell$ with different times at the start of reionization,
 $z_{\rm ion}=7$ and $z_{\rm ion}=9$ in terms of red shift.
For $z_{\rm ion}=7$,
 the overall magnitude of the free electron density increases
 because of the constraint of eq.(\ref{eq-optical-depth}),
 and the low-$\ell$ magnitude of $D^{\rm BB}_\ell$ increases,
 instead of decreasing the high-$\ell$ magnitude of $D^{\rm BB}_\ell$,
 due to the shorter time period available to produce polarizations.
The opposite happens for $z_{\rm ion}=9$.
We shall need other observables to determine the time at the beginning of reionization,
 and to reduce the uncertainty of predictions.
\begin{figure}[t]
\centering
\includegraphics[width=50mm]{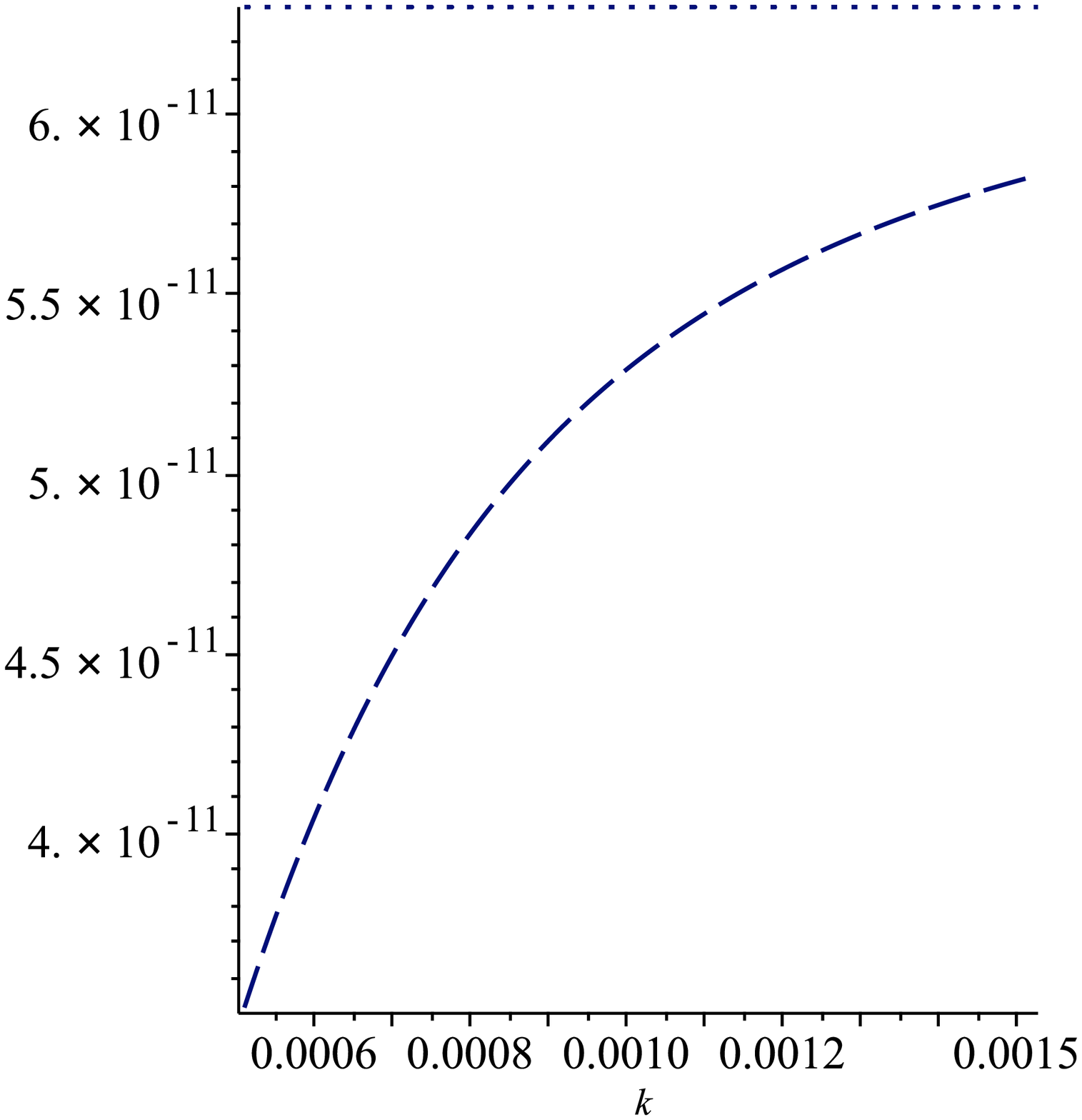}
\quad
\includegraphics[width=50mm]{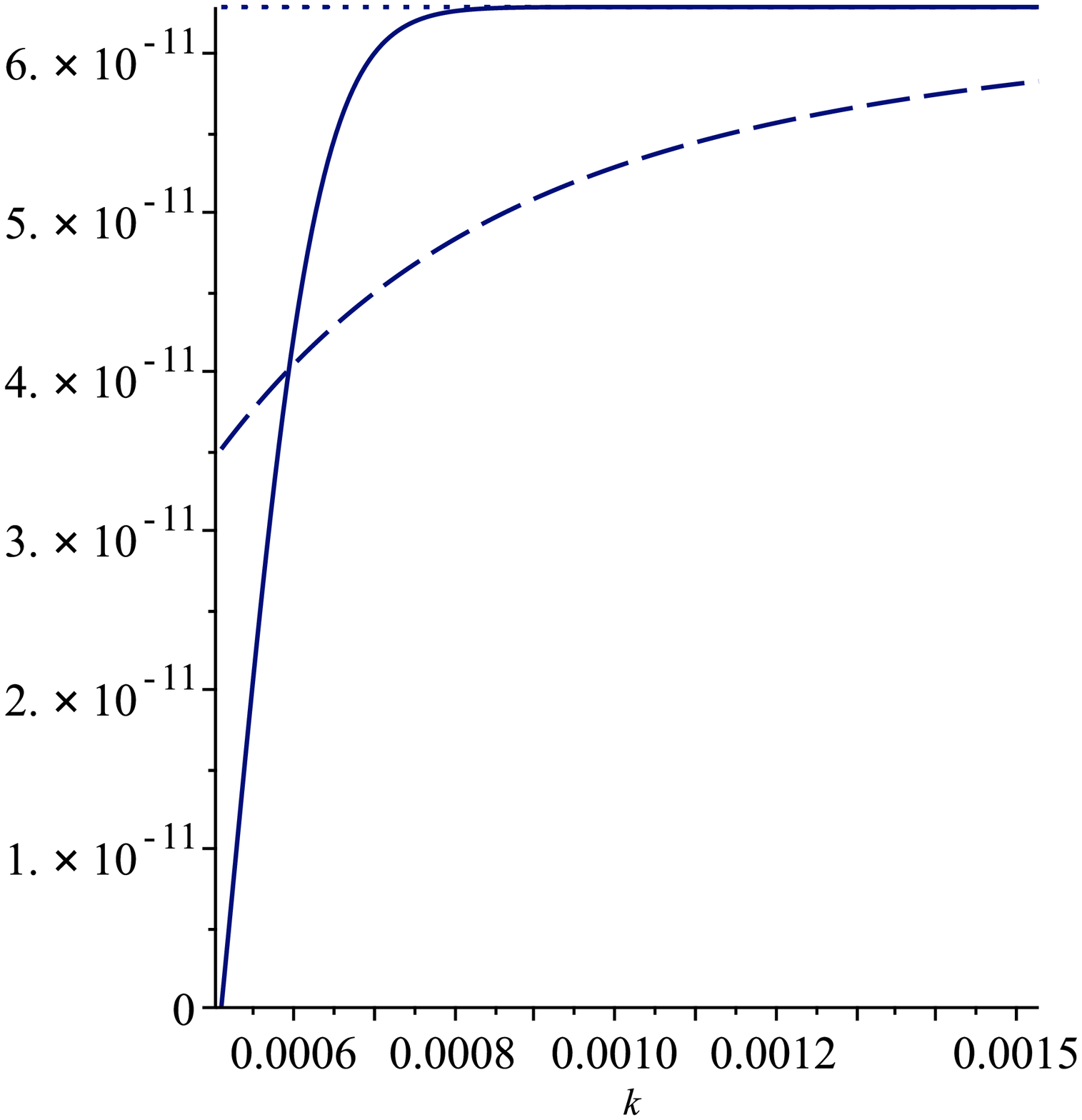}
\quad
\includegraphics[width=50mm]{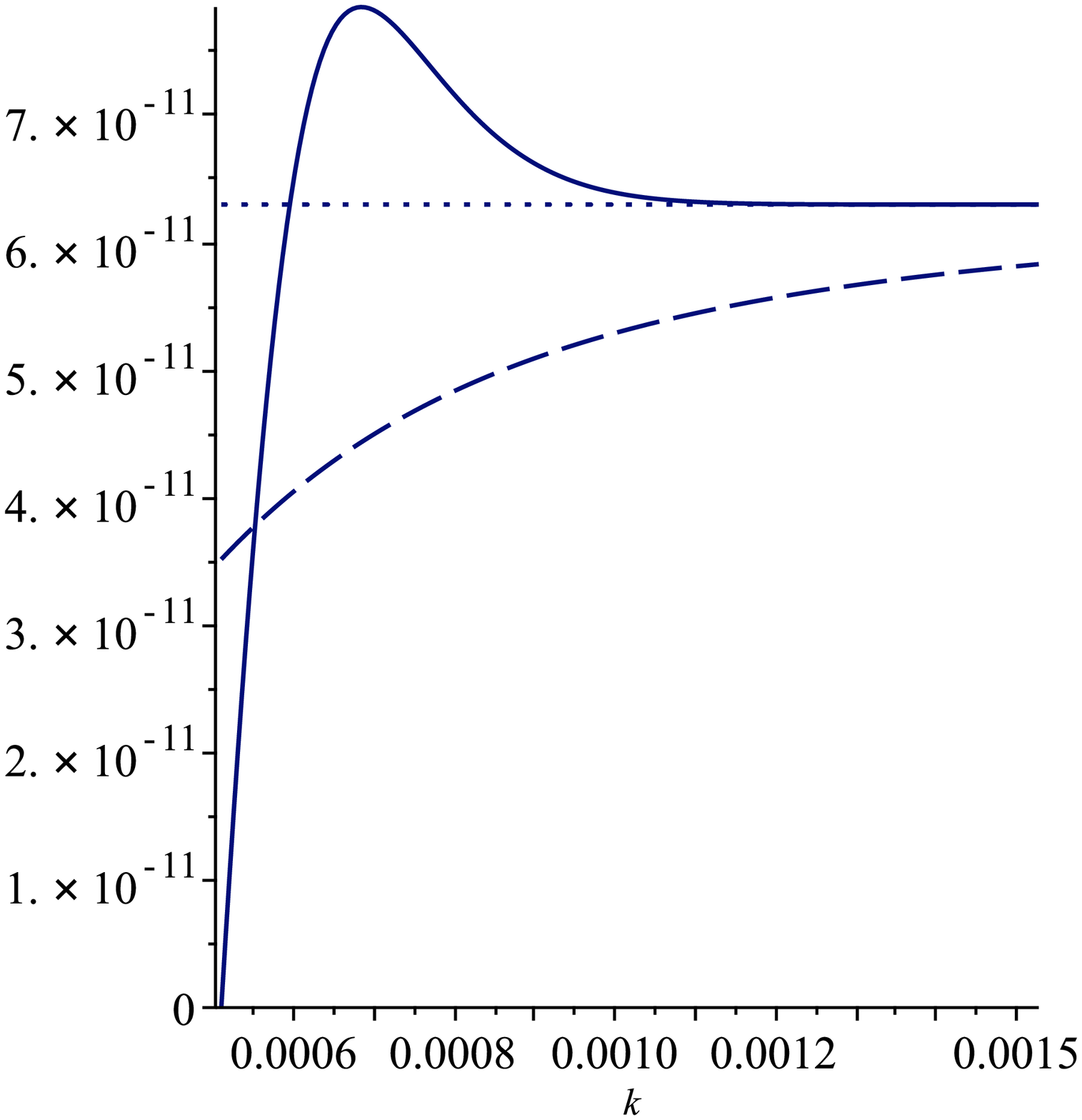}
\caption{
Three types of possible primordial tensor power spectra with $k$ in [$\rm{Mpc}^{-1}$],
 from left to right:
 for mild cut-off with $\Delta=0.351 \times 10^{-3}$ [$\rm{Mpc}^{-1}$],
 for sharp cut-off,
 and for sharp cut-off and ``overshoot'', or a peak.
The dotted line in each panel indicates the spectrum in $\Lambda$CDM model.
}
\label{fig:tensor-power}
\end{figure}
\begin{figure}[t]
\centering
\includegraphics[width=50mm]{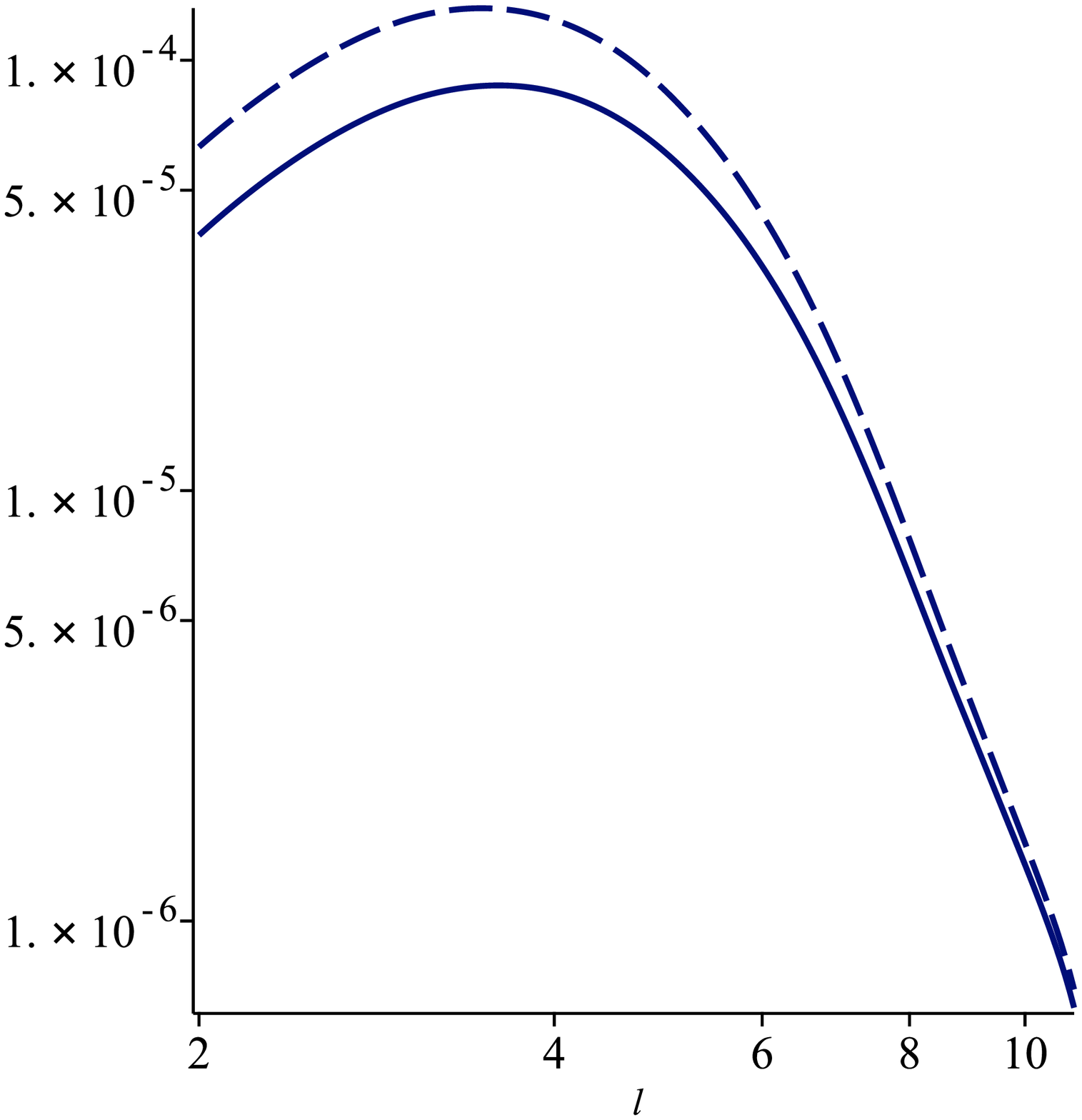}
\quad
\includegraphics[width=50mm]{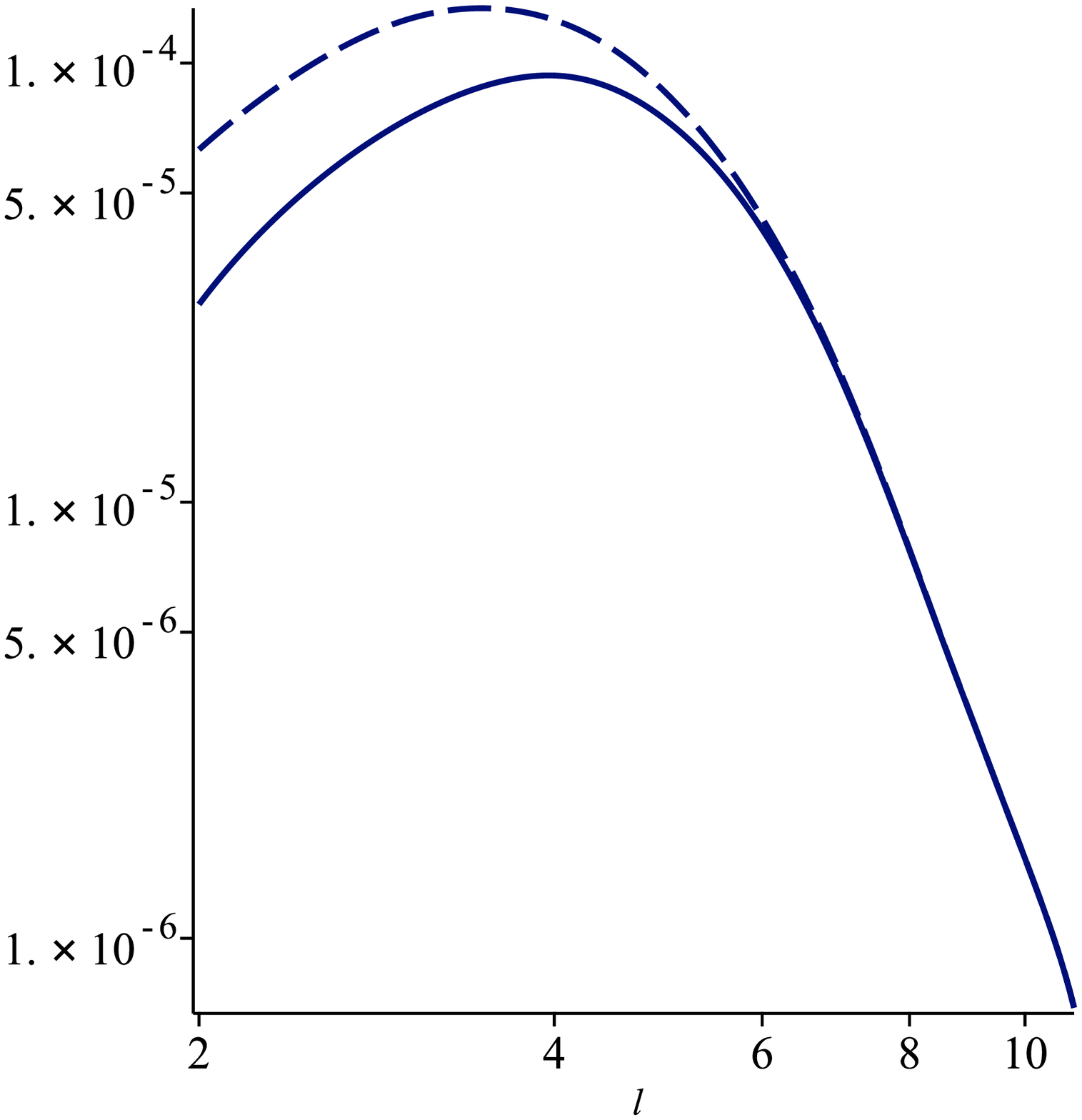}
\quad
\includegraphics[width=50mm]{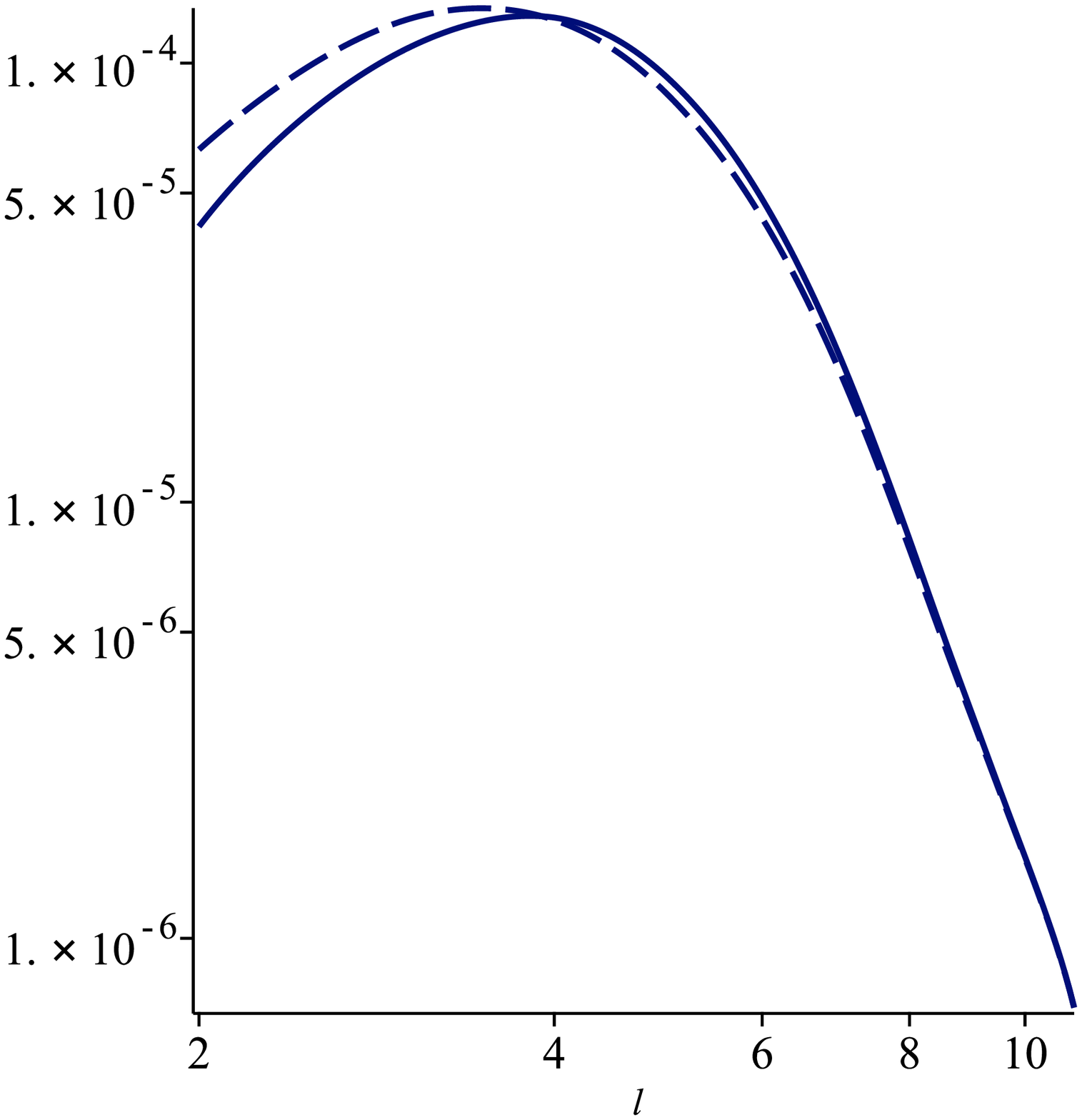}
\quad
\includegraphics[width=50mm]{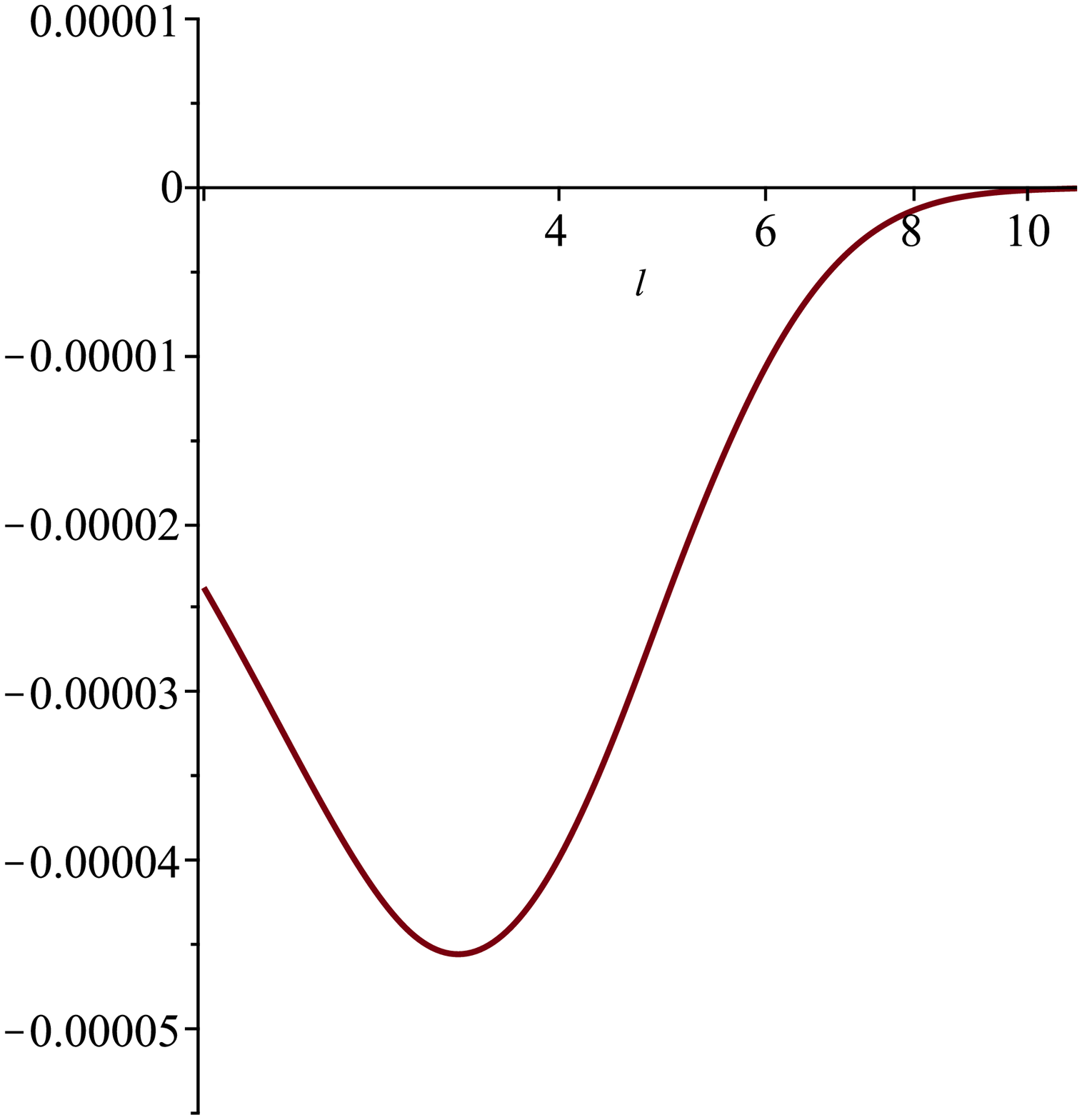}
\quad
\includegraphics[width=50mm]{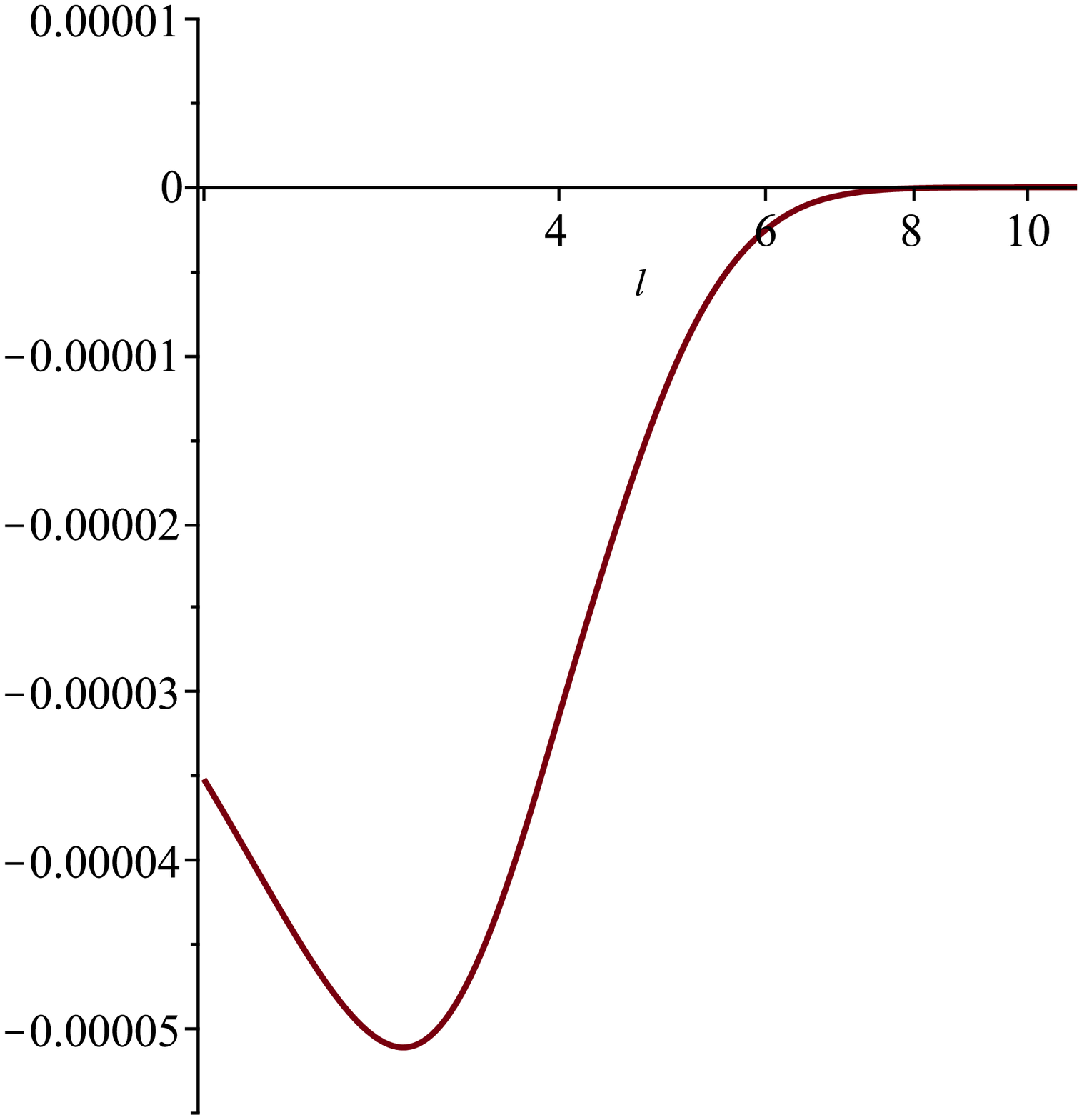}
\quad
\includegraphics[width=50mm]{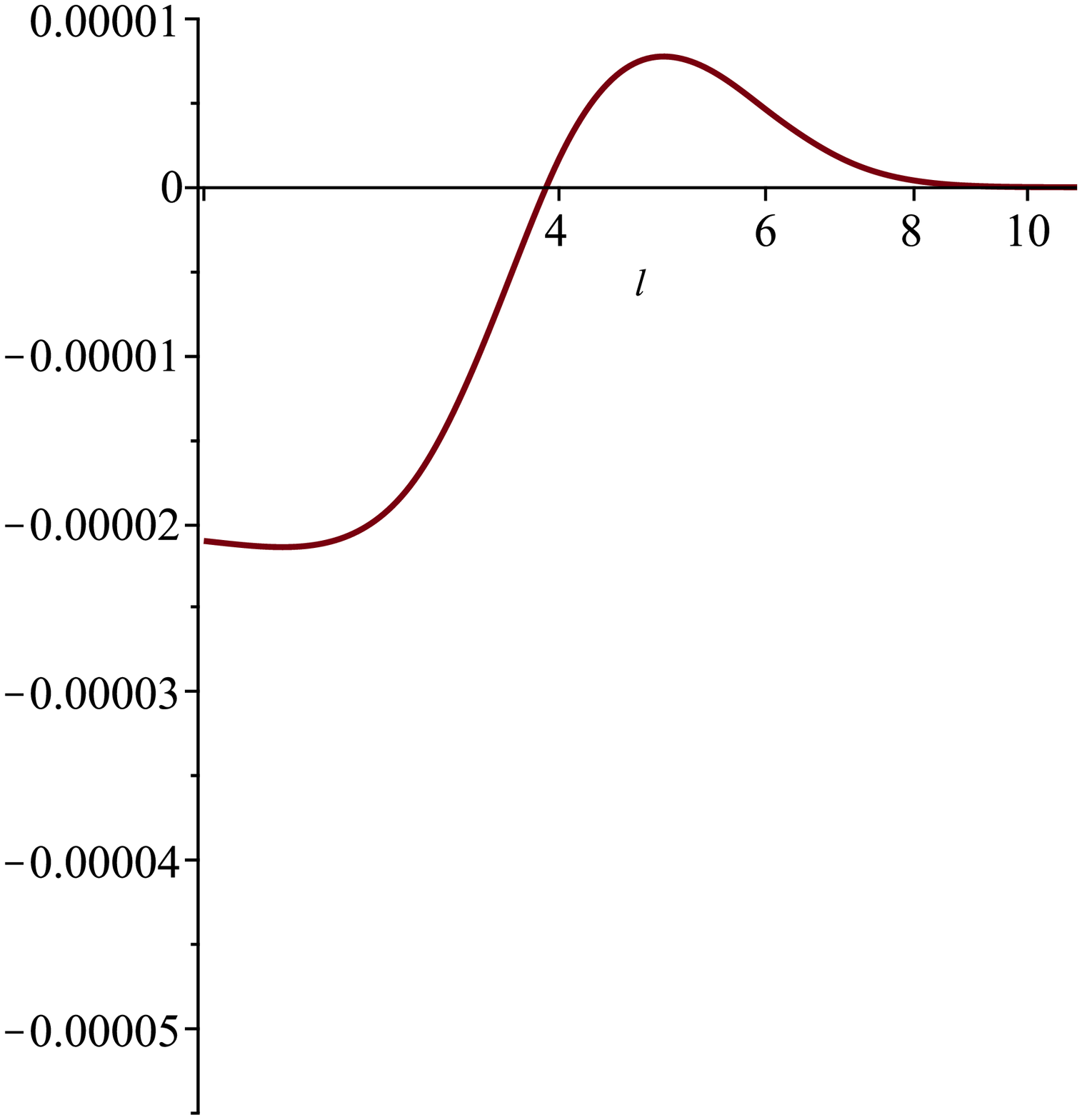}
\caption{
Three plots in upper line show
$D^{\rm BB}_\ell = (\ell(\ell+1)/2\pi) C^{\rm BB}_\ell$ [$\mu {\rm K}^2$]
 for various primordial tensor power spectra, from left to right:
 for cut-off induced by $\Delta=0.351 \times 10^{-3}$ [$\rm{Mpc}^{-1}$],
 for sharp cut-off,
 and for sharp cut-off and ``overshoot'', or a peak.
The dashed line in each panel indicates the prediction of $\Lambda$CDM.
Three plots in lower line show the differences with respect to the $\Lambda$CDM prediction,
 from left to right:
 for cut-off induced by $\Delta=0.351 \times 10^{-3}$ [$\rm{Mpc}^{-1}$],
 for sharp cut-off,
 and for sharp cut-off and ``overshoot'', or a peak.
}
\label{fig:D^BB_ell-inst-various}
\end{figure}

Let us now consider the effects of deformations of the primordial tensor power spectrum
 with instantaneous reionization at $z_{\rm ion}=8$.
Three types of possible primordial tensor power spectra are displayed in Fig.\ref{fig:tensor-power}.
The left panel shows the power spectrum of eq.(\ref{primordial-tensor-Delta}) with $n_T = 0$
\begin{equation}
 P^T_k = A_T \frac{k^3}{(k^2+\Delta^2)^{3/2}},
\end{equation}
 namely the spectrum with the typical cutoff,
 which is supported by present observations of temperature perturbations
 as a solution of the lack of power at low-$\ell$.
This captures the essence, but not the details, for example,
 of the climbing scenario \cite{Dudas:2012vv,Kitazawa:2014dya},
 since it does not account for a pre--inflationary peak.
The middle panel displays a power spectrum with a sharper cut-off
\begin{equation}
 P^T_k = A_T \tanh\left( \frac{k^2}{\Delta^2} - \frac{k_{\rm min}^2}{\Delta^2} \right),
\end{equation}
 where $k_{\rm min} \equiv 2 \pi/\eta_0$,
 and the right panel displays a similar power spectrum with a sharper cut-off and an ``overshoot''
 that is meant to simulate the typical pre--inflationary peaks of~\cite{Kitazawa:2014dya}.
There are simple analytic expressions for these types of spectra.
Aside from the family that can be found in~\cite{Dudas:2012vv}, one can also use
\begin{equation}
 P^T_k = A_T \left\{
              \tanh \left( \frac{k^2}{\Delta^2} - \frac{k_{\rm min}^2}{\Delta^2} \right)
              + \frac{k^2-k_{\rm min}^2}{\Delta^2} e^{-(k^2-k_{\rm min}^2)/\Delta^2}
             \right\} \ .
\end{equation}
Such a overshoot is typical of the transition from fast-roll era to slow-roll in the absence of a bounce
 (see \cite{Destri:2009hn}, for example).
Note that
 the last two power spectra are merely reproducing of the gross features of these types of power spectra,
 but are not solutions of specific models~\footnote{
A.Sagnotti and A.Gruppuso used similar expressions
 to investigate the effect of realistic pre--inflationary peaks on temperature spectra,
 with no significant improvements with respect to~\cite{Gruppuso:2015zia} and \cite{Gruppuso:2017nap}.
This is consistent with the failure to detect the second parameter $\gamma$
 in the first of these papers (private communication).
}.
Corresponding B-mode power spectra $D^{\rm BB}_\ell$
 are given in fig.\ref{fig:D^BB_ell-inst-various} in order, respectively.
These power spectra are easily understood by eq.(\ref{eq-beta-master}) (and eq.(\ref{eq-C^BB}))
 considering the modifications of $D_k$ as functions of $k$.
\begin{figure}[t]
\centering
\includegraphics[width=40mm]{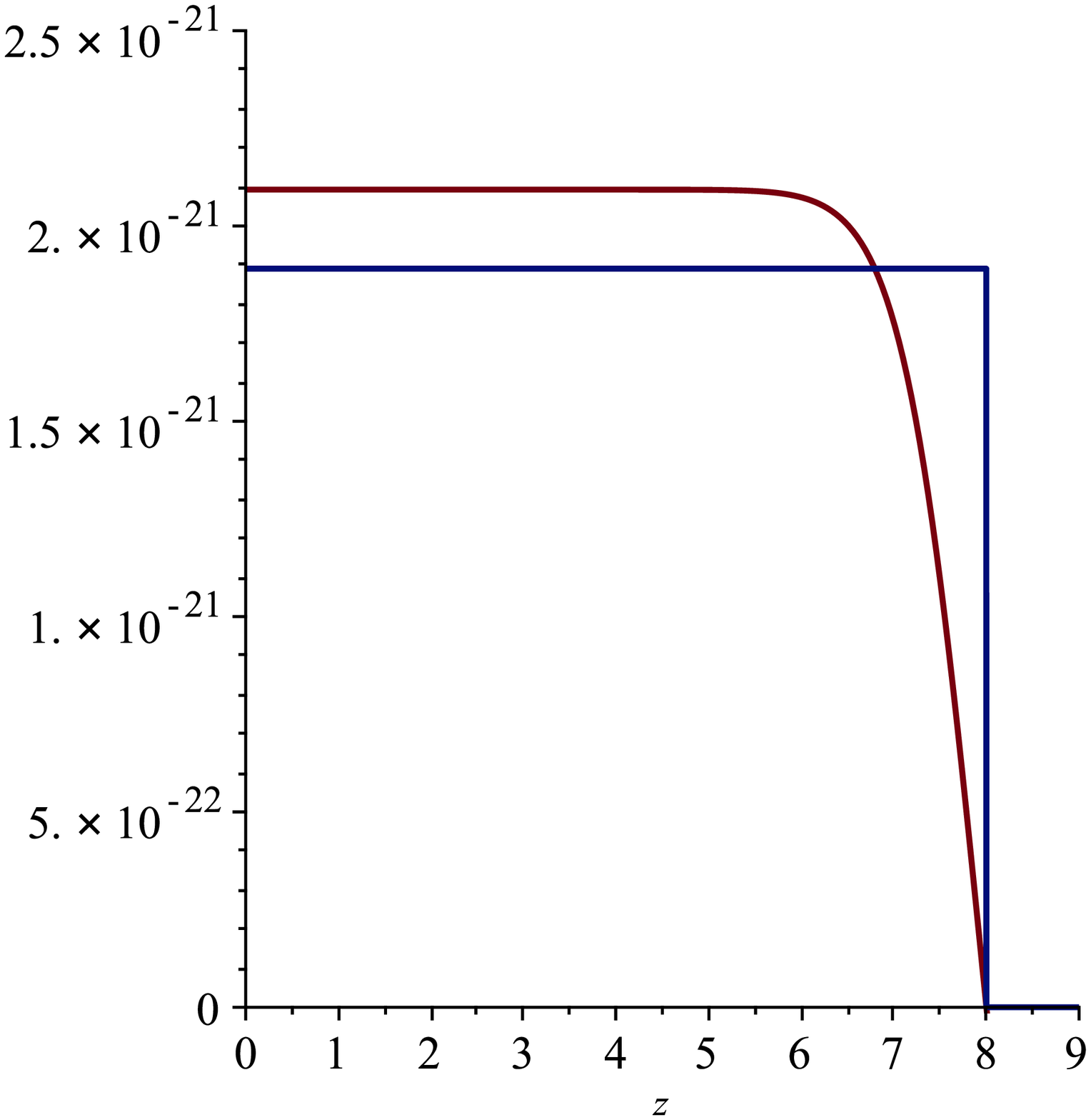}
\includegraphics[width=40mm]{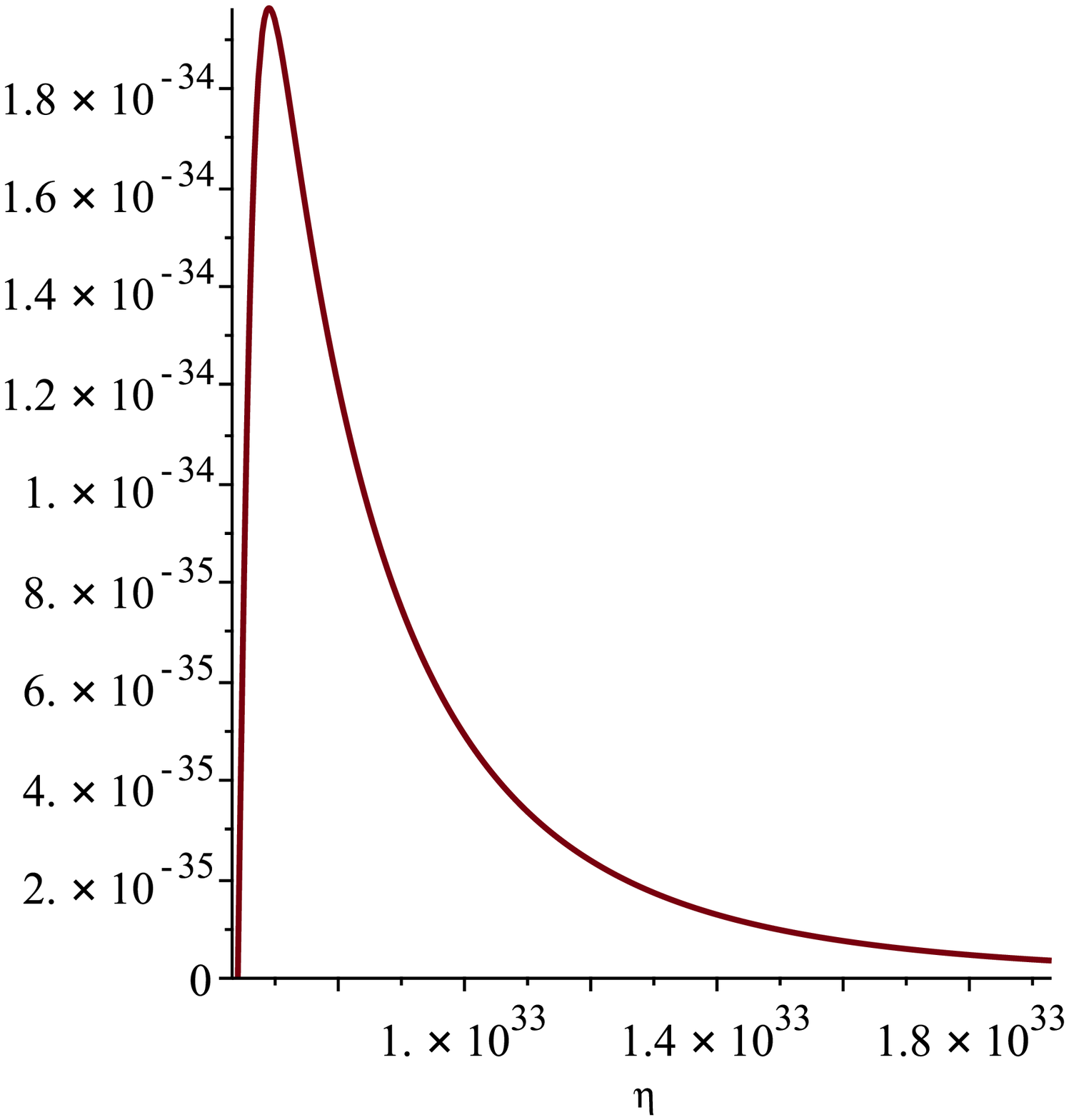}
\includegraphics[width=40mm]{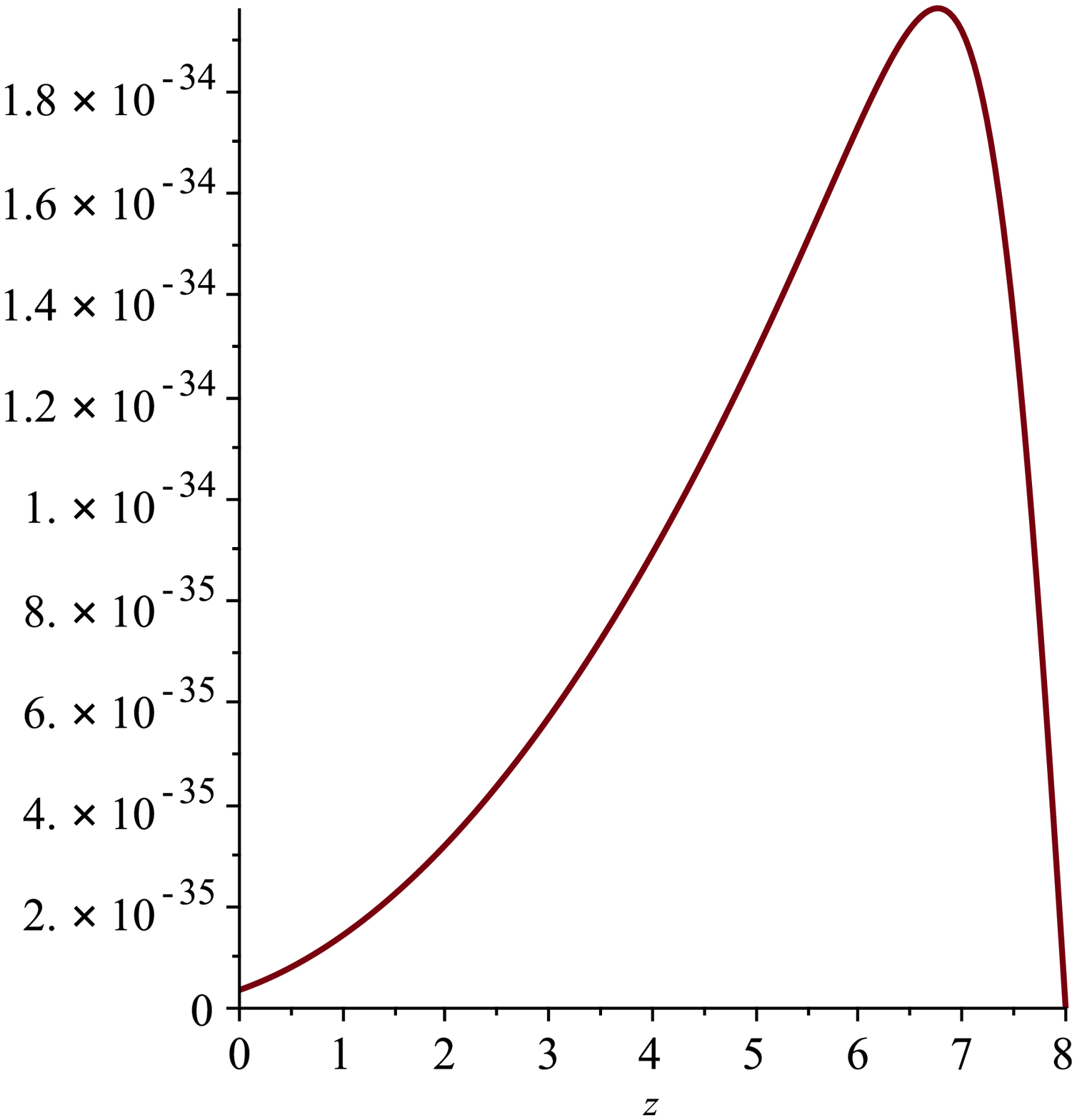}
\includegraphics[width=40mm]{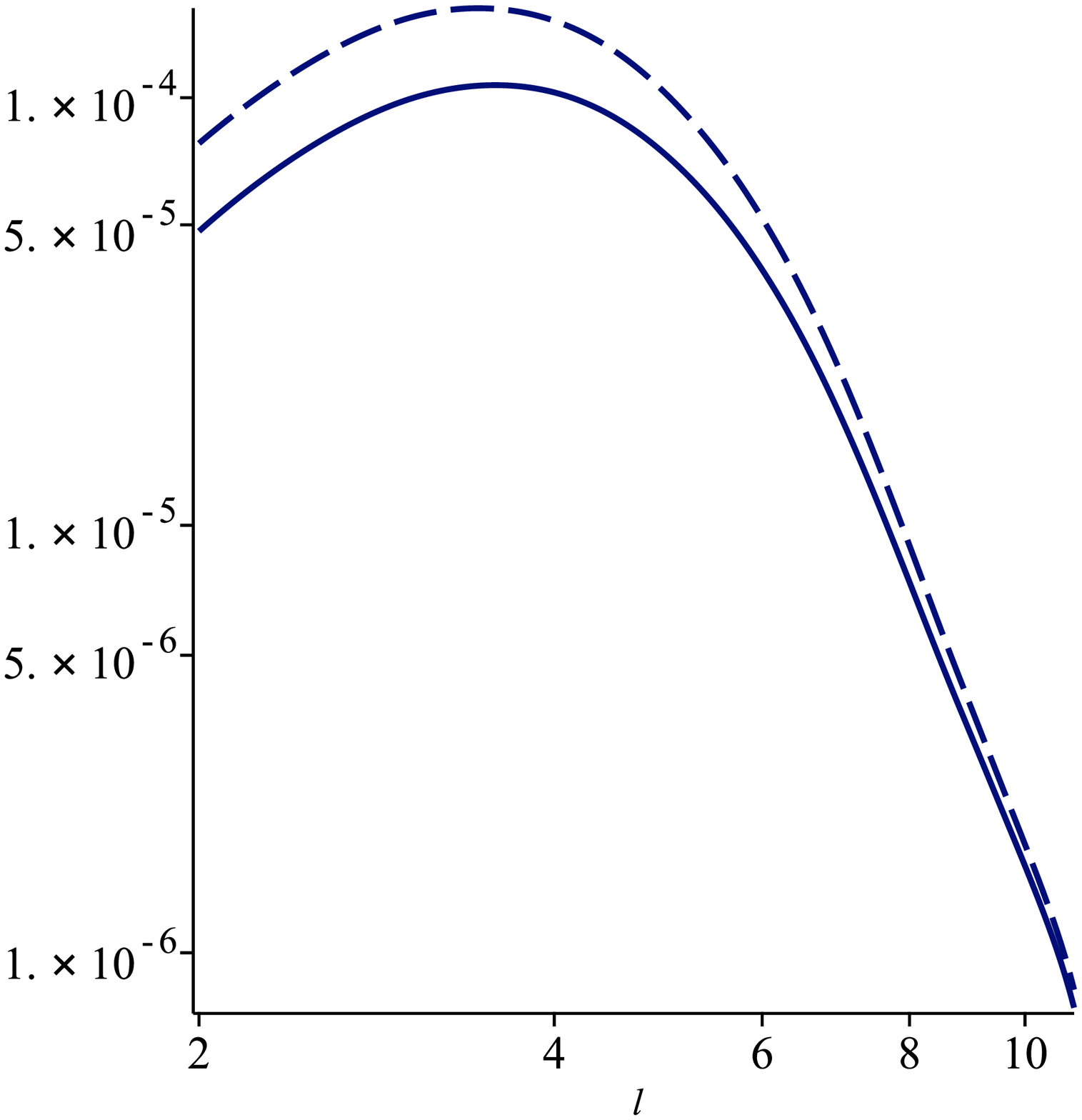}
\caption{
Left:
$n_e(z)a(z)^3$ [$\rm{eV}^3$] for the case of slow increase of electron density (red)
 and instantaneous reionization (blue). 
Middle-left:
$g$ [$\rm{eV}$] as a function of conformal time $\eta$ [$\rm{eV}^{-1}$]
 in the case of slow increase of the free electron density.
Middle-right:
$g$ [$\rm{eV}$] as a function of redshift $z$
 in the case of slow increase of the free electron density.
Right:
$D^{\rm BB}_\ell = (\ell(\ell+1)/2\pi) C^{\rm BB}_\ell$ [$\mu {\rm K}^2$]
 for $\Delta=0$ (dashed line) and $\Delta=0.351 \times 10^{-3}$ [$\rm{Mpc}^{-1}$] (solid line)
 in the case of slow increase of the free electron density.
}
\label{fig:g-real}
\end{figure}

For the standard cut-off with $\Delta$,
 the suppression of B-mode polarization power occurs for all values of $\ell$,
 as it is shown in the left panel of fig.\ref{fig:D^BB_ell-inst-various}.
 Presumably, one should be able to observe it with future probes, like LiteBIRD,
 despite the large uncertainty due to the cosmic variance.
We will return to the statistical significance of this case in the next section.
Here, we should remind our assumption, which realizes an ideal situation, that
 the B-mode polarization by primordial tensor perturbation is dominant
 in the region of $\ell \lesssim 10$, and the other contributions,
 mainly the B-mode polarization by scalar perturbations through gravitational lensing effect,
 are negligible or subtracted.
For the sharper cut-off,
 as shown in the middle panel of fig.\ref{fig:D^BB_ell-inst-various},
 the suppression typically concerns only the first few multipoles, $\ell = 2,3,4,5$,
 where the effect of the cosmic variance is large,
 and the observation will be difficult, if not impossible, even with future probes.
Of course, if the threshold of the cut were at larger value of $k$, the situation would be different,
 but the consistency with the observed low-$\ell$ temperature power spectrum would then be problematic.
Furthermore,
 if there were an overshoot in the primordial tensor power spectrum,
 as discussed in~\cite{Destri:2009hn} or in~\cite{Dudas:2012vv},
 detecting it would be more difficult,
 as one can see in the right panel of fig.\ref{fig:D^BB_ell-inst-various}.

Finally, let us consider the dependence on the reionization process,
considering a slow increase of the free electron density.
The left panel of fig.~\ref{fig:g-real}
 shows the time evolution of the free electron density,
 excluding the effect of the expansion of the Universe.
Rather than the step function, which corresponds to instantaneous reionization,
 we consider a slow increase of the free electron density modelled by a $\tanh$ function.
According to the observation of the Gunn-Peterson trough,
we choose it so that it increases about 90\% of the full amount until $z=6$.
The middle panel of fig.\ref{fig:g-real} shows the shape of $g(\eta)$,
 which should be compared to the left panel of fig.\ref{fig:g-and-beta}.
The B-mode polarization power spectrum $D^{\rm BB}_\ell$,
 with $\Delta=0.351 \times 10^{-3}$ [$\rm{Mpc}^{-1}$],
 is displayed in the right panel of fig.\ref{fig:g-real}.
It is almost the same as in the case of instantaneous reionization,
 but the overall magnitude is slightly enhanced,
 because of the enhancement of the free electron density at late times.
Again,
 the suppression of B-mode polarization power spectrum
 occurs for all values of $\ell$ in the relevant range.
Since the time evolution of free electron density in the process of reionization
 has been strongly constrained,
 the effect on the B-mode polarization power spectrum is limited.
The dependence on the red shift values of beginning the reionization is more important,
 as we have already seen.

\section{Conclusions}
\label{sec:conclusions}

In this paper we have proposed a simple semi--analytic method
 to investigate the low-$\ell$ B-mode polarization power spectrum
 $D^{\rm BB}_\ell = (\ell(\ell+1)/2\pi) C^{\rm BB}_\ell$.
The method rests on a truncation, which affects its accuracy,
 but has the virtue of leading to the expressions in eqs.(\ref{eq-beta-master}) and (\ref{eq-C^BB})
 which are simple enough to highlight the Physics underlying the emergence of B-mode polarization.
Moreover,
 the predictions of the $\Lambda$CDM model derived in this fashion
 are consistent with the results of more precise numerical calculations,
 even in a certain quantitative level.
We have also investigated
 how the incomplete knowledge of the reionization process affects the B-mode spectrum,
 identifying the uncertainty on the start of reionization as the largest source of uncertainty.

\begin{figure}[t]
\centering
\includegraphics[width=50mm]{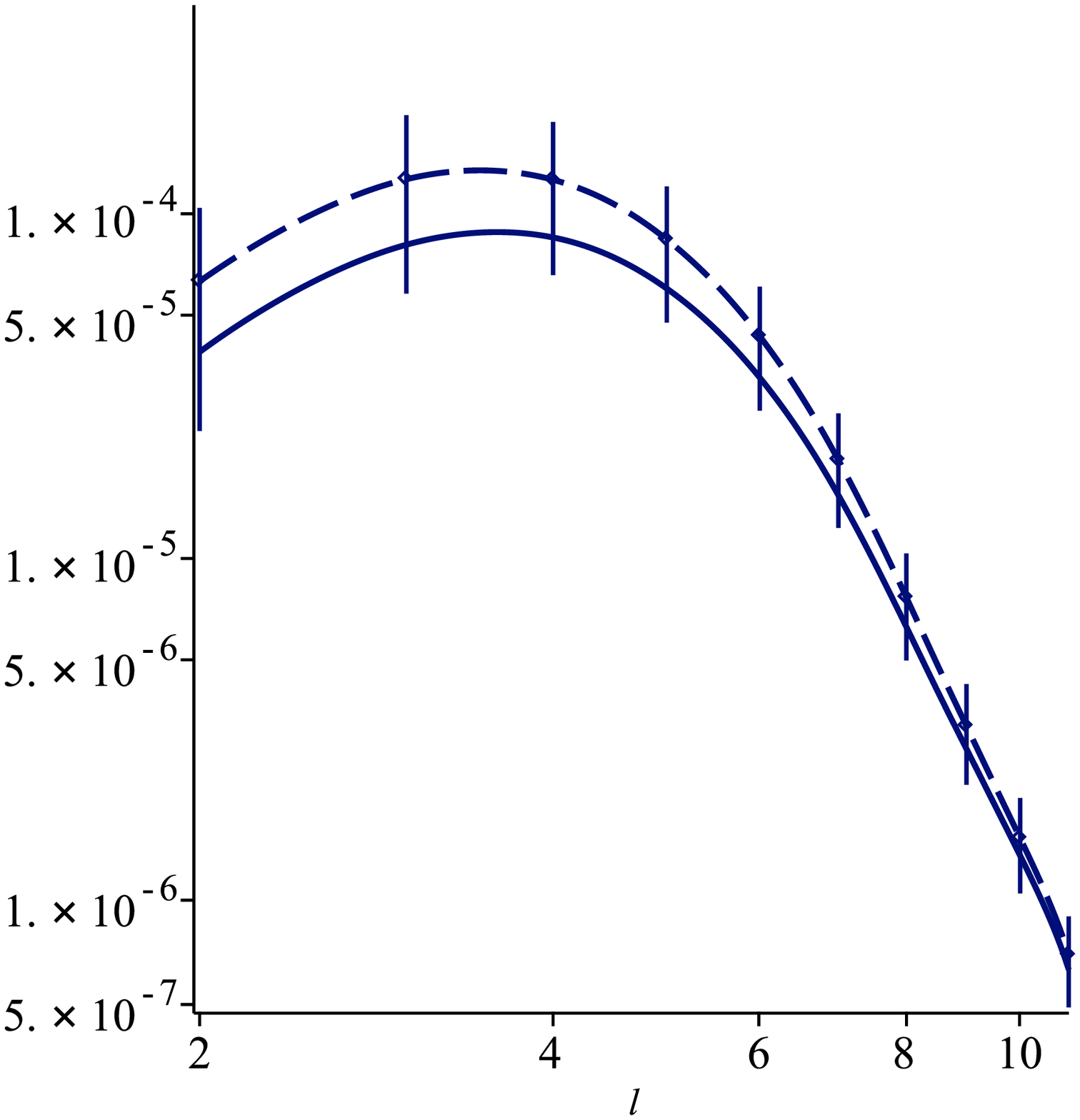}
\includegraphics[width=50mm]{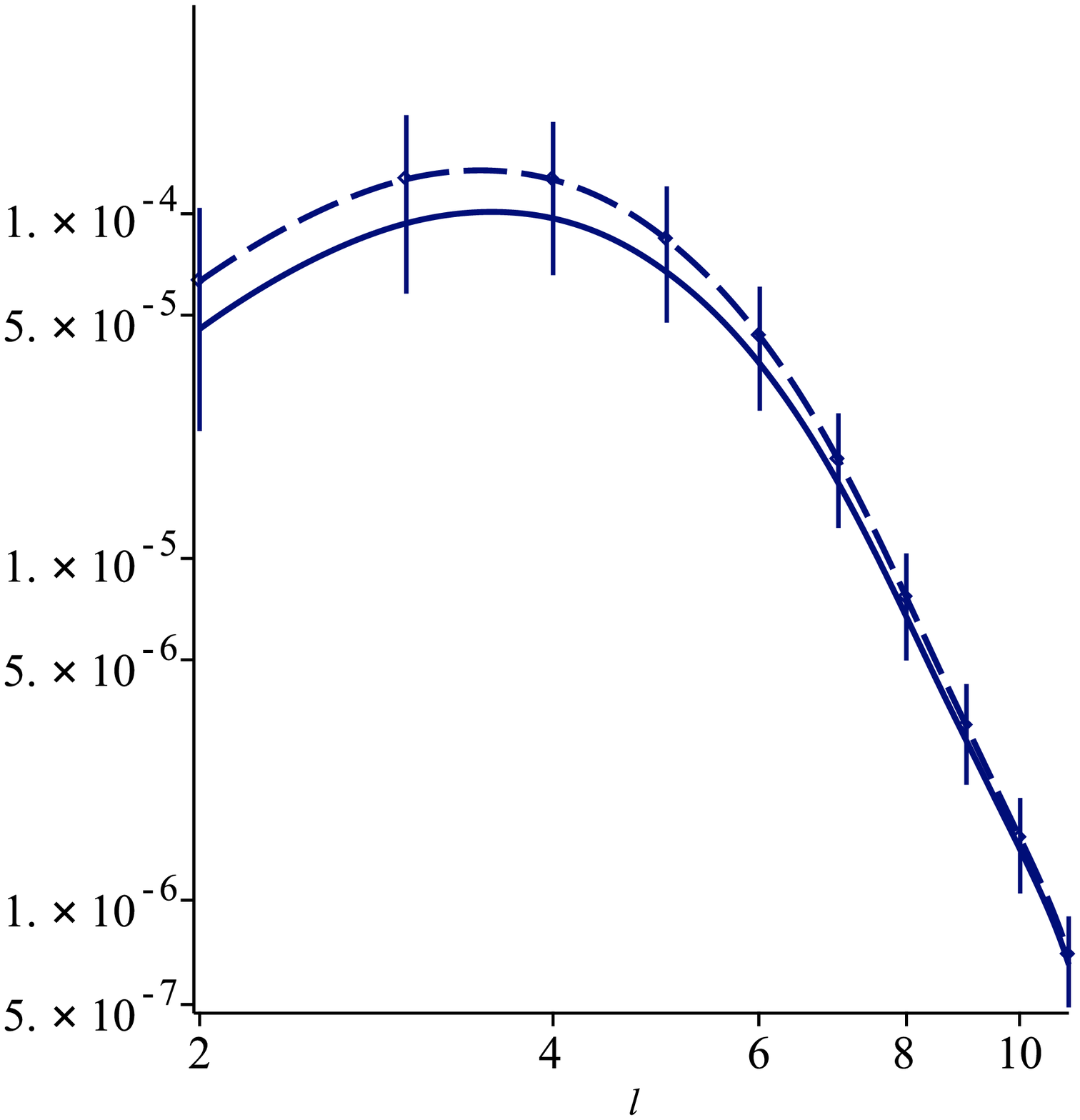}
\includegraphics[width=50mm]{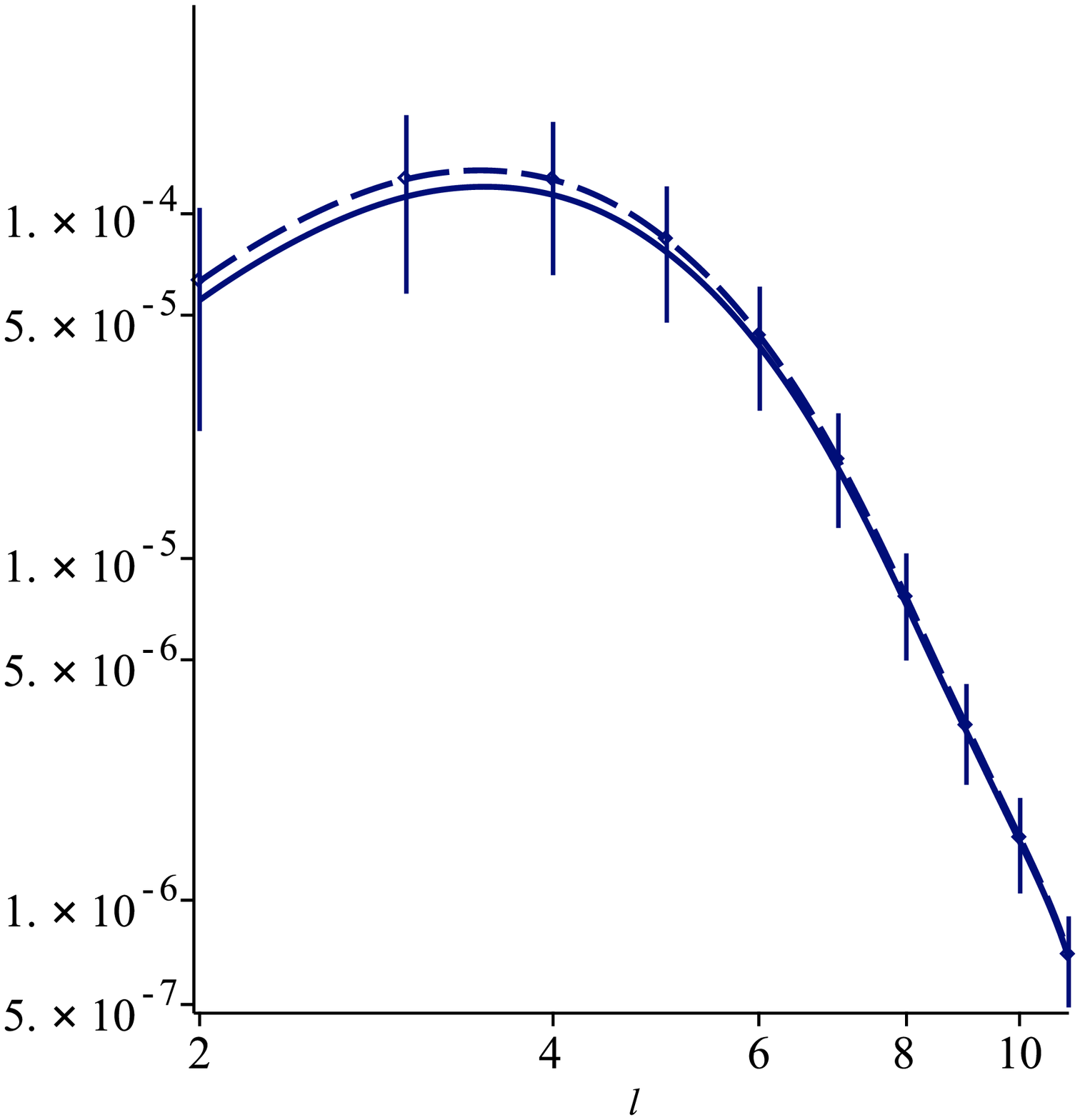}
\caption{
$D^{\rm BB}_\ell = (\ell(\ell+1)/2\pi) C^{\rm BB}_\ell$ [$\mu {\rm K}^2$]
 with the uncertainties due to the cosmic variance.
Dashed lines indicate the prediction of the $\Lambda$CDM model
 with uncertainties due to the cosmic variance,
 and solid lines indicate the predictions
 for the primordial tensor power spectrum of eq.~(\ref{primordial-tensor-Delta})
 with $\Delta = 0.351 \times 10^{-3}$ [$\rm{Mpc}^{-1}$] (left panel),
 $\Delta = 0.28 \times 10^{-3}$ [$\rm{Mpc}^{-1}$] (middle panel), and
 $\Delta = 0.17 \times 10^{-3}$ [$\rm{Mpc}^{-1}$] (right panel).
}
\label{fig:D^BB_ell-with-error}
\end{figure}

Our main theme
 has been the effect of possible modifications of the primordial tensor power spectrum at large scales.
The deformed power spectrum of eq.~(\ref{primordial-tensor-Delta})
 yields the systematic power reduction for low-$\ell$ shown in fig.\ref{fig:D^BB_ell-with-error},
 where the uncertainties due to cosmic variance, estimated as
 $\sqrt{2/(2\ell+1)} \times 100$\% of the value for each $\ell$,
 are included as error bars.
All in all, however, our results depart slightly from the $\Lambda$CDM model,
 since the $p$-value,
 the probability that an observation provide a worse fit with the $\Lambda$CDM prediction
 than eq.~(\ref{primordial-tensor-Delta})
 with the value $\Delta=0.351 \times 10^{-3}$ [$\rm{Mpc}^{-1}$]
 of~\cite{Gruppuso:2015xqa,Gruppuso:2017nap}, is $p=0.98$.

\begin{figure}[t]
\centering
\includegraphics[width=50mm]{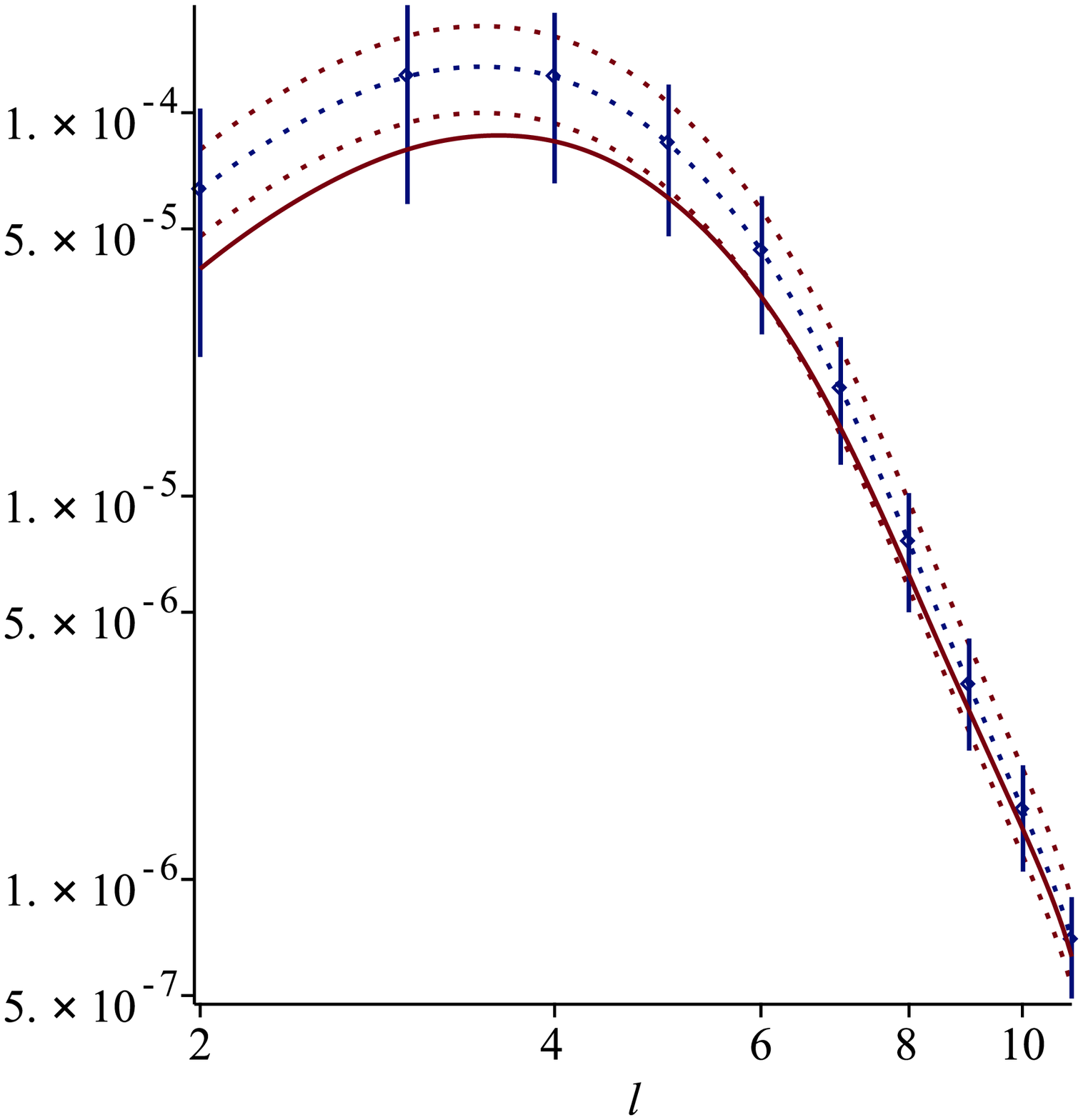}
\includegraphics[width=50mm]{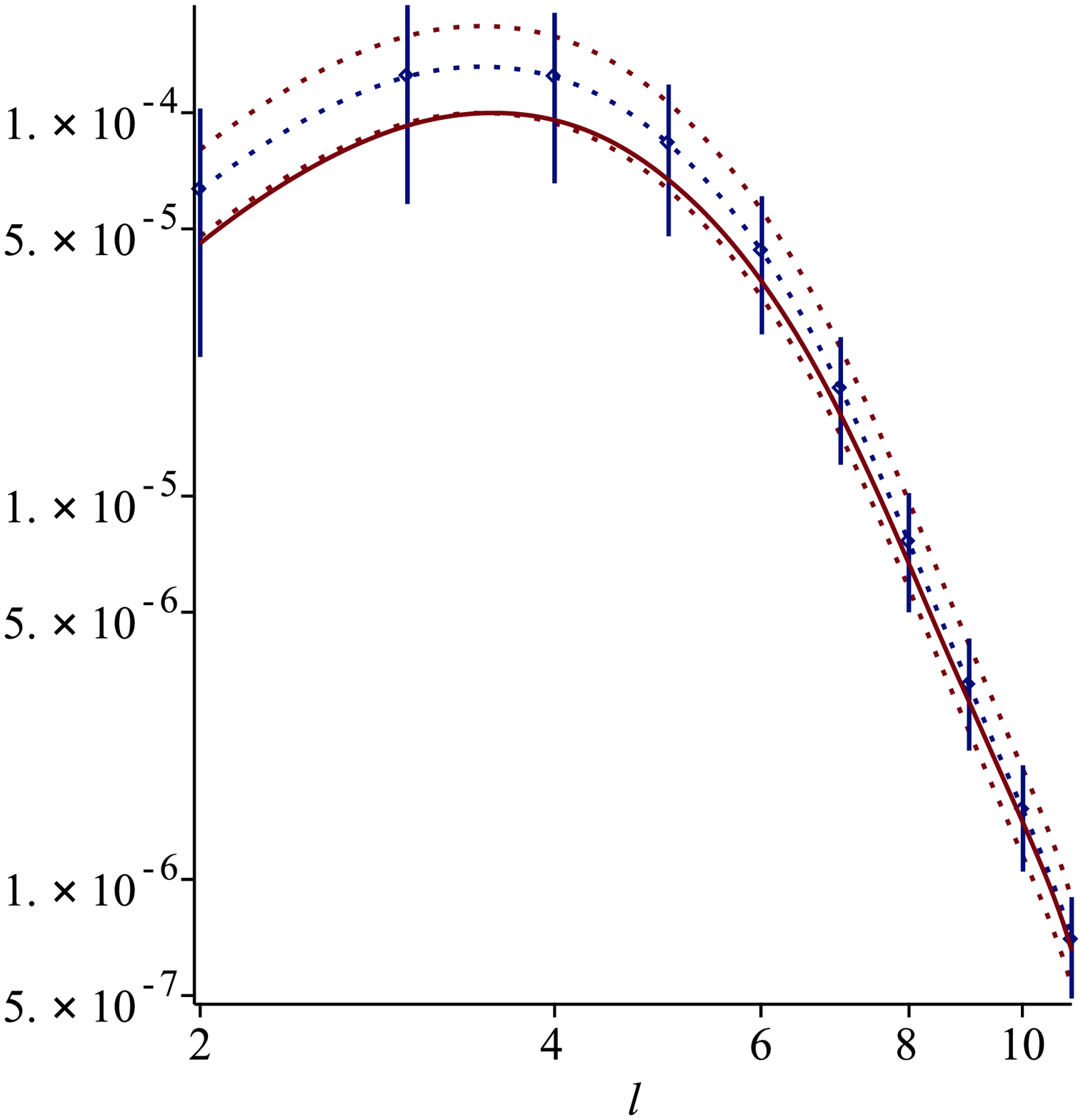}
\includegraphics[width=50mm]{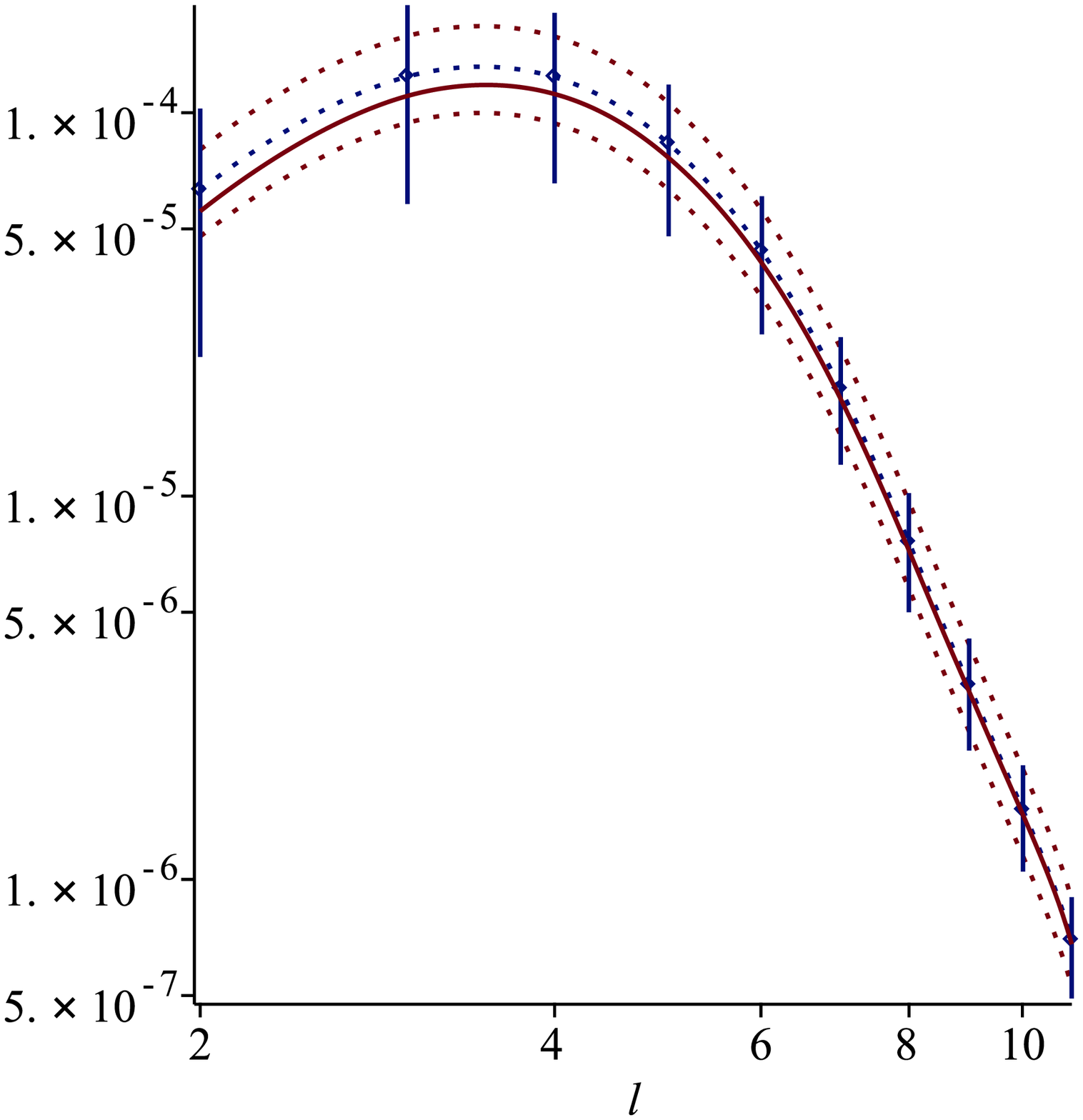}
\caption{
$D^{\rm BB}_\ell = (\ell(\ell+1)/2\pi) C^{\rm BB}_\ell$ [$\mu {\rm K}^2$]
 with the uncertainties due to the cosmic variance and optical depth $\tau$.
For each panel
 the dashed line with error bars from cosmic variance
 indicates the prediction of the $\Lambda$CDM model with the central value of $\tau = 0.054 \pm 0.007$
 and the dashed lines above and below indicate those of possible larger and smaller vales of $\tau$
 within $1\sigma$.
The solid lines in each panel have the same meaning as in fig.\ref{fig:D^BB_ell-with-error}.
}
\label{fig:D^BB_ell-with-tau-error}
\end{figure}

This statistical analysis, however,
 does not take into account the fact that the $\Lambda$CDM$\Delta$ model predicts polarizations
 that are {\it systematically smaller} than those of $\Lambda$CDM model.
A better handle would require some finer statistical probe,
 for instance the variance (the auto-correlation function up to $\ell_{\rm max}$)
\begin{equation}
 V(\ell_{\rm max})
  = \sum_{\ell=2}^{\ell_{\rm max}} \frac{2\ell+1}{4\pi} C^{BB}_\ell \ ,
\end{equation}
 which was introduced to investigate the lack of the power at low-$\ell$
 in temperature perturbations
 \cite{Monteserin:2007fv,Cruz:2010ud,Gruppuso:2013xba}.
In the present case, for B-mode polarizations with $\ell_{\rm max}=11$,
 $V_{\Lambda{\rm CDM}} = (1.2 \pm 0.30) \times 10^{-4}$ $\mu {\rm K}^2$
 and $V_{\Lambda{\rm CDM}\Delta} \simeq 0.78 \times 10^{-4}$ $\mu {\rm K}^2$, so that
 the prediction of the $\Lambda$CDM$\Delta$ model
 deviates by about $1.3\sigma$ from the predictions of $\Lambda$CDM model.

The value of $\Delta = 0.351 \times 10^{-3}$ [$\rm{Mpc}^{-1}$]
 was obtained from the fit of data with best statistical significance,
 using an extended galactic mask that grants a 39\% sky coverage \cite{Gruppuso:2015xqa,Gruppuso:2017nap}.
With smaller galactic masks, the value of $\Delta$ becomes smaller:
 $\Delta = 0.28 \times 10^{-3}$ [$\rm{Mpc}^{-1}$] granting a 59\% sky coverage, and
 $\Delta = 0.17 \times 10^{-3}$ [$\rm{Mpc}^{-1}$] with the standard galactic mask
 granting a 94\% sky coverage.
The corresponding $D^{\rm BB}_\ell$ for these smaller values of $\Delta$
 are shown in the middle and right panels of fig.\ref{fig:D^BB_ell-with-error}.
 The corresponding variances are
 $V_{\Lambda{\rm CDM}\Delta} \simeq 0.89 \times 10^{-4}$ [$\mu {\rm K}^2$]
 and
 $V_{\Lambda{\rm CDM}\Delta} \simeq 1.1 \times 10^{-4}$ [$\mu {\rm K}^2$],
 and are consistent with the predictions of the $\Lambda$CDM model
 (less than $1\sigma$ away from its prediction).
Fig.\ref{fig:D^BB_ell-with-tau-error} shows the plots of the same of fig.\ref{fig:D^BB_ell-with-error}
 with the uncertainty of the value of optical depth $\tau = 0.054 \pm 0.007$ \cite{Aghanim:2018eyx}.
The change of the value of $\tau$
 shifts the normalization of the spectrum with small $\ell$ dependence,
 which should be compared with the larger $\ell$ dependent distortions by larger values of $\Delta$.
The details of the reionization process
 can only have a limited impact on our considerations,
 insofar as the uncertainty on $\tau$ remains of the order of 10\%
\footnote{The author would like to thank A.Gruppuso and P.Natoli for raising this point.}.

The uncertainty of the value of reionization redshift $z_{\rm ion}$
 also affects the investigation of $\Delta$
 (see Figs.\ref{fig:D^BB_ell-inst} and \ref{fig:D^BB_ell-inst-various}).
The ambiguity of the value of $z_{\rm ion}$
 causes a little horizontal parallel shift of the B-mode angular power spectrum,
 which differs however from a larger vertical suppression induced by $\Delta$.
Therefore,
 it is fair to expect that the ambiguity of $z_{\rm ion}$
 should not be very important for the determination of $\Delta$.
Anyway,
 it will be important to include the precise knowledge about reionization process
 by future experiments, like Square Kilometer Array, for example.
The joint analysis of TT and EE or TT, EE and BB power spectra will be effective,
 since the interplays between $\tau$ and $\Delta$ are different in each power spectrum.
We leave this analysis for future work.

In conclusion,
 we have highlighted the main features of the phenomena
 at the origin of low--$\ell$ B-mode polarization with a simple method,
 but we are presently unable to gather a convincing evidence linking the lack of power
 that can be potentially observed in forthcoming experiments on B-mode polarization
 to the start of inflation.
More accurate numerical investigations
 will be necessary for a more precise assessment of the statistical significance of these effects.
This might also allow, in principle,
 to investigate further primordial tensor perturbations with different features
 around the time of the start of inflation.

\section*{Acknowledgments}

The author would like to thank A.~Gruppuso, P.~Natoli and A.~Sagnotti for helpful discussions
 and a careful reading of the manuscript.
The author would also like to thank Scuola Normale Superiore
 and the University of Ferrara for the kind hospitality.
This work was supported in part by Scuola Normale Superiore and by the JSPS KAKENHI Grant Number 19K03851.


\begin{thebibliography}{99}


\bibitem{Akrami:2018vks}
  Y.~Akrami {\it et al.} [Planck Collaboration],
  arXiv:1807.06205 [astro-ph.CO].

\bibitem{Hinshaw:1996ut}
  G.~Hinshaw, A.~J.~Banday, C.~L.~Bennett, K.~M.~Gorski, A.~Kogut, C.~H.~Lineweaver, G.~F.~Smoot
   and E.~L.~Wright,
  Astrophys.\ J.\  {\bf 464} (1996) L25
  [astro-ph/9601061].
\bibitem{Spergel:2003cb}
  D.~N.~Spergel {\it et al.} [WMAP Collaboration],
  Astrophys.\ J.\ Suppl.\  {\bf 148} (2003) 175
  [astro-ph/0302209].
\bibitem{Sarkar:2010yj}
  D.~Sarkar, D.~Huterer, C.~J.~Copi, G.~D.~Starkman and D.~J.~Schwarz,
  Astropart.\ Phys.\  {\bf 34} (2011) 591
  [arXiv:1004.3784 [astro-ph.CO]].
\bibitem{Gruppuso:2013dba}
  A.~Gruppuso,
  Mon.\ Not.\ Roy.\ Astron.\ Soc.\  {\bf 437} (2014) no.3,  2076
  [arXiv:1310.2822 [astro-ph.CO]].
\bibitem{Ade:2013nlj}
  P.~A.~R.~Ade {\it et al.} [Planck Collaboration],
  Astron.\ Astrophys.\  {\bf 571} (2014) A23
  [arXiv:1303.5083 [astro-ph.CO]].
\bibitem{Akrami:2018odb}
  Y.~Akrami {\it et al.} [Planck Collaboration],
  arXiv:1807.06211 [astro-ph.CO].

\bibitem{Gruppuso:2015zia}
  A.~Gruppuso and A.~Sagnotti,
  Int.\ J.\ Mod.\ Phys.\ D {\bf 24} (2015) no.12,  1544008
  [arXiv:1506.08093 [astro-ph.CO]].

\bibitem{chib_mukh}
V.~F.~Mukhanov and G.~V.~Chibisov,
  JETP Lett.\  {\bf 33} (1981) 532
   [Pisma Zh.\ Eksp.\ Teor.\ Fiz.\  {\bf 33} (1981) 549].

\bibitem{Aghanim:2018eyx}
  N.~Aghanim {\it et al.} [Planck Collaboration],
  arXiv:1807.06209 [astro-ph.CO].

\bibitem{Dimopoulos:2016yep}
  K.~Dimopoulos and M.~Artymowski,
  Astropart.\ Phys.\  {\bf 94} (2017) 11
  [arXiv:1610.06192 [astro-ph.CO]].

\bibitem{Destri:2009hn}
  C.~Destri, H.~J.~de Vega and N.~G.~Sanchez,
  Phys.\ Rev.\ D {\bf 81} (2010) 063520
  [arXiv:0912.2994 [astro-ph.CO]].
\bibitem{Dudas:2012vv}
  E.~Dudas, N.~Kitazawa, S.~P.~Patil and A.~Sagnotti,
  JCAP {\bf 1205} (2012) 012
  [arXiv:1202.6630 [hep-th]].
\bibitem{Kitazawa:2014dya}
  N.~Kitazawa and A.~Sagnotti,
  JCAP {\bf 1404} (2014) 017
  [arXiv:1402.1418 [hep-th]].


\bibitem{Gruppuso:2015xqa}
  A.~Gruppuso, N.~Kitazawa, N.~Mandolesi, P.~Natoli and A.~Sagnotti,
  Phys.\ Dark Univ.\  {\bf 11} (2016) 68
  [arXiv:1508.00411 [astro-ph.CO]].
\bibitem{Gruppuso:2017nap}
  A.~Gruppuso, N.~Kitazawa, M.~Lattanzi, N.~Mandolesi, P.~Natoli and A.~Sagnotti,
  Phys.\ Dark Univ.\  {\bf 20} (2018) 49
  [arXiv:1712.03288 [astro-ph.CO]].

\bibitem{bsb}
  S.~Sugimoto,
  Prog.\ Theor.\ Phys.\  {\bf 102} (1999) 685 [arXiv:hep-th/9905159];
  I.~Antoniadis, E.~Dudas and A.~Sagnotti,
  Phys.\ Lett.\ {\bf B 464} (1999) 38 [arXiv:hep-th/9908023];
  C.~Angelantonj,
  Nucl.\ Phys.\ {\bf B 566} (2000) 126 [arXiv:hep-th/9908064];
  G.~Aldazabal and A.~M.~Uranga,
  JHEP {\bf 9910} (1999) 024 [arXiv:hep-th/9908072];
  C.~Angelantonj, I.~Antoniadis, G.~D'Appollonio, E.~Dudas and A.~Sagnotti,
  Nucl.\ Phys.\ {\bf B 572} (2000) 36 [arXiv:hep-th/9911081].

\bibitem{stringtheory}
  For reviews see: M.~B.~Green, J.~H.~Schwarz and E.~Witten, ``Superstring Theory'', 2 vols.,
  Cambridge, UK: Cambridge Univ. Press (1987);
  J.~Polchinski, ``String theory'', 2 vols. Cambridge, UK: Cambridge Univ. Press (1998);
  C.~V.~Johnson, ``D-branes,'' USA: Cambridge Univ. Press (2003) 548 p;
  B.~Zwiebach, ``A first course in string theory'' Cambridge, UK: Cambridge Univ. Press (2004);
  K.~Becker, M.~Becker and J.~H.~Schwarz,
   ``String theory and M-theory: A modern introduction'' Cambridge, UK: Cambridge Univ. Press (2007);
  E.~Kiritsis, ``String theory in a nutshell'', Princeton, NJ: Princeton Univ. Press (2007).

\bibitem{Angelantonj:2002ct}
  C.~Angelantonj and A.~Sagnotti,
  Phys.\ Rept.\  {\bf 371} (2002) 1
   Erratum: [Phys.\ Rept.\  {\bf 376} (2003) no.6,  407]
  [hep-th/0204089];
J.~Mourad and A.~Sagnotti,
  arXiv:1711.11494 [hep-th].

\bibitem{climbing}
  E.~Dudas, N.~Kitazawa and A.~Sagnotti,
  Phys.\ Lett.\ B {\bf 694} (2010) 80
  [arXiv:1009.0874 [hep-th]];
  A.~Sagnotti,
  Phys.\ Part.\ Nucl.\ Lett.\  {\bf 11} (2014) 836
  [arXiv:1303.6685 [hep-th]];
  P.~Fr\'e, A.~Sagnotti and A.~S.~Sorin,
  Nucl.\ Phys.\ {\bf B 877} (2013) 1028
  [arXiv:1307.1910 [hep-th]].

\bibitem{Zaldarriaga:1998ar}
  M.~Zaldarriaga and U.~Seljak,
  Phys.\ Rev.\ D {\bf 58} (1998) 023003
  [astro-ph/9803150].

\bibitem{Polnarev:1985}
  A.G.~Polnarev,
  Sov.\ Astron.\ {\bf 29} (1985) 607.


\bibitem{Mukhanov:2003xr}
  V.~F.~Mukhanov,
  Int.\ J.\ Theor.\ Phys.\  {\bf 43} (2004) 623
  [astro-ph/0303072].
\bibitem{Mukhanov:2005sc}
  V.~Mukhanov,
  ``Physical Foundations of Cosmology,''
  Cambridge, UK: Univ. Pr. (2005).

\bibitem{Cabella:2004mk}
  P.~Cabella and M.~Kamionkowski,
  astro-ph/0403392.
\bibitem{Keating:1997cv}
  B.~Keating, P.~Timbie, A.~Polnarev and J.~Steinberger,
  Astrophys.\ J.\  {\bf 495} (1998) 580
  [astro-ph/9710087].
\bibitem{Pritchard:2004qp}
  J.~R.~Pritchard and M.~Kamionkowski,
  Annals Phys.\  {\bf 318} (2005) 2
  [astro-ph/0412581].
\bibitem{Zhang:2005nv}
  W.~Zhao and Y.~Zhang,
  Phys.\ Rev.\ D {\bf 74} (2006) 083006
  [astro-ph/0508345].

\bibitem{Basko:2080}
  M.~M.~Basko and A.~G.~Polnarev,
  Mon.\ Not.\ Roy.\ Astron.\ Soc.\  {\bf 191} (2080) 207.

\bibitem{Weinberg:2003ur}
  S.~Weinberg,
  Phys.\ Rev.\ D {\bf 69} (2004) 023503
  [astro-ph/0306304].

\bibitem{Zaldarriaga:1996xe}
  M.~Zaldarriaga and U.~Seljak,
  Phys.\ Rev.\ D {\bf 55} (1997) 1830
  [astro-ph/9609170].

\bibitem{Zaldarriaga:1995gi}
  M.~Zaldarriaga and D.~D.~Harari,
  Phys.\ Rev.\ D {\bf 52} (1995) 3276
  [astro-ph/9504085].


\bibitem{Adam:2016hgk}
  R.~Adam {\it et al.} [Planck Collaboration],
  Astron.\ Astrophys.\  {\bf 596} (2016) A108
  [arXiv:1605.03507 [astro-ph.CO]].

\bibitem{Gunn:1965hd}
  J.~E.~Gunn and B.~A.~Peterson,
  Astrophys.\ J.\  {\bf 142} (1965) 1633.

\bibitem{Ishino:2016wxl}
  H.~Ishino,
  Int.\ J.\ Mod.\ Phys.\ Conf.\ Ser.\  {\bf 43} (2016) 1660192.


\bibitem{Monteserin:2007fv}
  C.~Monteserin, R.~B.~B.~Barreiro, P.~Vielva, E.~Martinez-Gonzalez, M.~P.~Hobson and A.~N.~Lasenby,
  Mon.\ Not.\ Roy.\ Astron.\ Soc.\  {\bf 387} (2008) 209
  [arXiv:0706.4289 [astro-ph]].
\bibitem{Cruz:2010ud}
  M.~Cruz, P.~Vielva, E.~Martinez-Gonzalez and R.~B.~Barreiro,
  Mon.\ Not.\ Roy.\ Astron.\ Soc.\  {\bf 412} (2011) 2383
  [arXiv:1005.1264 [astro-ph.CO]].
\bibitem{Gruppuso:2013xba}
  A.~Gruppuso, P.~Natoli, F.~Paci, F.~Finelli, D.~Molinari, A.~De Rosa and N.~Mandolesi,
  JCAP {\bf 1307} (2013) 047
  [arXiv:1304.5493 [astro-ph.CO]].

\end{thebibliography}
\end{document}